\documentclass[12pt, a4paper]{article}
\usepackage{latexsym,amsfonts,amssymb}
\usepackage{graphicx}
\usepackage{indentfirst}
\usepackage{cite}
\usepackage[cp1251]{inputenc}

\newtheorem{theorem}{Theorem}
\newtheorem{lema}{Lemma}

\newtheorem{definition}{Definition}

\newtheorem{remark}{Remark}

\begin{document}

\begin{center}
{\Large \bf Lie symmetries and reductions of multi-dimensional  boundary value problems of the Stefan type}

\medskip

{\bf Roman Cherniha$^{\dag,}$ $^\ddag$} {\bf and  Sergii Kovalenko$^\dag$}
\\
{\it $^\dag$~Institute of Mathematics, Ukrainian National Academy of Sciences,
\\
3 Tereshchenkivs'ka Street, Kyiv 01601, Ukraine}
\\
{\it $^\ddag$~Department  of  Mathematics, National University `Kyiv-Mohyla Academy',
\\
2 Skovoroda Street, Kyiv  04070 ,  Ukraine}
\\

\medskip
 E-mail: cherniha@imath.kiev.ua and kovalenko@imath.kiev.ua
\end{center}

\begin{abstract}
A new definition of Lie invariance for  nonlinear multi-dimensional   boundary value problems (BVPs) is  proposed  by the  generalization of known definitions to much wider classes of  BVPs. The class of  (1+3)-dimensional nonlinear BVPs of the Stefan type, modeling the process of melting and evaporation of metals, is studied in detail. Using the definition proposed,
the group classification  problem  for  this class of BVPs is solved   and  some  reductions (with physical meaning)  to BVPs of
lower dimensionality   are made. Examples  of how to construct
exact solutions of the  (1+3)-dimensional nonlinear BVP with the
correctly-specified coefficients  are presented.

\medskip

\end{abstract}

%2010 Mathematics Subject Classification : 22E70, 35K61, 80A22.

%PACS numbers: 02.20.Qs, 02.20.Sv, 02.30.Jr, 02.60.Lj, 64.70.D-, 64.70.F-,
%     72.15.Eb.

\section{\bf Introduction}

Currently, Lie symmetries are widely applied
to study partial differential equations (PDEs) (including multi-component systems
of multi-dimensional PDEs), notably, for  their reductions to ordinary differential equations (ODEs) and constructing  exact solutions.
There are a vast number of papers and many excellent
books (see, e.g.,  \cite{bl-anco02, b-k, fss, olv, ovs} and references
cited therein) devoted to such applications. However,  one may note that
the authors usually do not pay any attention to the application
of  Lie symmetries for  solving  boundary
value problems (BVPs).  To the best  of our knowledge,
the first papers that did so were published at the beginning
of 1970s (see  \cite{pukh-72} and \cite{bl-1974} and  their  extended versions
 presented in books \cite{pukh-et-al-98} and
\cite{b-k}, respectively). The  books, which highlight essential
role of Lie symmetries in solving BVPs  and
present several examples, were published much later
\cite{b-k, rog-ames-89, ibr96}.

 The main object of this paper is a class of  (1+3)-dimensional  nonlinear BVPs of the Stefan type.
These  problems    are widely used  in
mathematical  modeling of a wide range of processes, which   arise in
physics,  biology, chemistry   and  industry \cite{alex93, bri-03, crank84,
ready, rub71, anisimov70}. Nevertheless, these processes can be very different
from the formal point of view, they have  the common peculiarity,
unknown moving boundaries (free boundaries). Movement of unknown boundaries is described by the famous Stefan boundary conditions \cite{st, rub71,  gupta}. It is well known that exact solutions of BVPs of the Stefan type  can be
derived only in exceptional cases and the relevant list is not very long
 at the present time (see \cite{alex93, ch-od90, ch93, br-tr02, br-tr10, br-tr07, voller10, barry08,ch-kov-09} and references cited therein). Notably, those   exact solutions were constructed under additional conditions on their form and/or the coefficients arising in  the relevant BVP.
 It should also be stressed that  all analytical results derived in those papers are concerned with two-dimensional BVPs. To the best of our knowledge,  there  are no  invariant  solutions (with physical meaning) of multidimensional  BVPs  with free boundaries excepting particular problems with radial symmetry, for which analytical solutions are found (see, e.g., \cite{alex93, crank75} and references cited therein). Perhaps \cite{pukh-06} (see  also  references by the same author cited therein) is unique because it contains such exact solutions for  (1+3)-dimensional hydrodynamical problems  with free boundaries.
 Of course, there are many interesting  papers, devoted to the rigorous asymptotic analysis of such BVPs,  leading to the relevant  analytical results (see,e.g., \cite{kartashov-01, king-et-all-03, king-et-all-05} and references cited therein).

From the mathematical point of view,
BVPs with free  boundaries are   more complicated
objects than the standard BVPs with fixed boundaries.
In a particular case, each  BVP with Stefan boundary conditions
is nonlinear; nevertheless, the basic equations may be linear \cite{rub71, cr-jg59}.
 Thus, the classical methods of solving linear BVPs (the Fourier method, the Laplace transformation and so forth)  cannot be directly applied
 for solving  any BVP with free  boundaries.
On the other hand,  it can be
noted that the Lie symmetry method  can  be more applicable just
for solving problems with moving boundaries than for other BVPs. In fact, the structure
of  unknown  boundaries may depend  on invariant variable(s) and this
provides the possibility to reduce the BVP in question  to that of lover
dimensionality. This is the reason why  some   authors have applied
the Lie symmetry method to BVPs with free  boundaries \cite{ bl-1974,
pukh-72, ben-olv-82, pukh-06, ch-2003, ch-kov-09, ch-kov-11}.

The paper is organized as follows.
  In  section 2, we  propose a new  definition of Lie invariance
  for any BVPs with basic evolution equations, which    generalize
  the  known definitions, and formulate the algorithm for solving
  the group classification problem.
    In  section 3,
we apply the definition and the algorithm to the class of (1+3)-dimensional
BVPs  of the Stefan type, used to describe melting and evaporation of
materials in the case when  their surface is exposed to a powerful
flux of energy. The  main result  is presented in Theorem 2, which  is a  highly non-trivial
 generalization of  that, derived  for  two-dimensional
 BVPs  in  \cite{ch-kov-09} .
 In section 4, all possible   systems of  subalgebras (optimal systems of subalgebras)
   for   a subclass of  (1+3)-dimensional BVPs  admitting  a five-dimensional  Lie algebra of invariance are constructed.
 We  reduce such
problems   to the two-dimensional  BVPs      via  the non-conjugate two-dimensional subalgebra.
Moreover, we show that this reduction admits a clear physical interpretation.    Examples of how
to construct  exact solutions   of the  (1+3)-dimensional nonlinear BVP with
the specified coefficients of diffusion   are presented too.
Finally, we present conclusions  in section~5.

\section{\bf Definition of Lie invariance for a BVP with free boundaries}

Consider a BVP for a  system of $n$ evolution equations ($n \geq 2$)
with $m + 1$ independent $(t, x)$ (hereafter $x = (x_1, x_2, \ldots,
x_m)$) and $n$ dependent $u = (u_1, u_2, \ldots, u_n)$ variables.
Let us assume that the basic equations possess the form
\begin{equation}\label{1}
u_t^i=F^i \left(x, u, u_x, \ldots , u_{x}^{(k_i)}\right), \ i = 1,
\ldots, n
\end{equation}
\noindent and are  defined on a domain $(0,+\infty)\times{\Omega}
\subset \mathbb{R}^{m + 1} $, where ${\Omega}$ is an open domain
with  smooth boundaries. Hereafter, the lower subscripts $t$ and $x$
denote differentiation with respect to these variables and
$u_{x}^{(k_i)}$ denotes a totality of  partial derivatives  with
respect to $x$ of order $k_i$ (for example, $u_{x}^{(1)}=u_{x_1},
\ldots, u_{x_m}$).

 Consider three types of boundary and initial conditions,
which widely arise in applications:
\begin{equation}\label{2}
s_a(t,x)=0: \ B^{j}_a \left(t,x, u, u_x, \ldots ,
u_{x}^{(k_{a}^j)}\right) = 0,\ a = 1, \ldots, p, \, j =1,\ldots,n_a,
\end{equation}
\begin{equation}\label{3}
S_b(t,x)=0: \ B^{l}_b \left(t,x, u, \ldots , u_{x}^{(k_b^l)}, S_b^{(1)},
\ldots , S^{(K^l_b)}_b \right) = 0, \ b = 1, \ldots, q, \, l
=1,\ldots,n_b,
\end{equation}
\noindent and
\begin{equation}\label{4}
\gamma_c(t,x)=\infty: \ \Gamma^{m}_c \left(t,x, u, u_x, \ldots ,
u_{x}^{(k_{c}^m)}\right) = 0, \ c = 1, \ldots, r, \, m = 1, \ldots,
n_c.
\end{equation}
\noindent Here $k_{a}^j < k=\max\{k_1,\ldots,k_n\}, \ k_{b}^l< k, \
k_{c}^m < k$ and $K^l_b$ are the given numbers, $s_a(t,x)$ and
$\gamma_c(t,x)$ are the known functions, while the functions $S_b(t,x)$
defining free boundary surfaces must be found. In (\ref{3}), the
function  $S^{(K^l_b)}_b$ denotes a totality of derivatives with
respect to $t$ and $x$ of order $K^l_b$ (for instance, $S_b^{(1)} = \frac{\partial S_b}{\partial t}, \frac{\partial S_b}{\partial x_1}, \ldots, \frac{\partial S_b}{\partial x_m}$).   We also assume that  all
functions arising in (\ref{1})--(\ref{4}) are sufficiently smooth so
that a classical solution exists for this BVP.

We note  that  boundary  conditions (\ref{4})  essentially differ from those    (\ref{2}) because they  are defined on the non-regular manifolds ${\cal{M}}_c = \{ \gamma_c(t,x)=\infty \}$. Such conditions appears if one considers BVPs  in the unbounded domains and often leads to difficulties. For example, one may check that  the definition of BVP invariance presented in \cite{b-k, bl-anco02} is not valid for such conditions.

Consider an $N$--parameter  (local) Lie group  $G_N$ of point
transformations of variables $(t,x,u)$ in the Euclidean space
$\mathbb{R}^{n+m+1}$ (open subset of $\mathbb{R}^{n+m+1}$) defined
by the equations
\begin{equation}\label{5}
t^{\ast} = T(t,x,\varepsilon), \ \ x^{\ast}_i = X_i
(t,x,\varepsilon), \ \  u^{\ast}_j = U_j(t,x,u,\varepsilon), \ i =
1, \ldots, m, \ j = 1, \ldots, n,
\end{equation}
\noindent where $\varepsilon = (\varepsilon_1, \varepsilon_2,
\ldots, \varepsilon_N)$ are the group parameters. According to the
general Lie group theory, one may construct the  corresponding
$N$-dimensional Lie algebra $L_N$ with the basic  generators
\begin{equation}\label{6}
X_\alpha = \xi^0_\alpha \frac{\partial}{\partial t}+\xi^1_\alpha
\frac{\partial}{\partial x_1} + \ldots + \xi^m_\alpha
\frac{\partial}{\partial x_m} + \eta^1_\alpha
\frac{\partial}{\partial u_1}+ \ldots +\eta^n_\alpha
\frac{\partial}{\partial u_n}, \ \alpha = 1,2, \ldots, N,
\end{equation}
\noindent where
%\begin{equation}
$\xi^0_\alpha = \left. \frac{\partial T(t,x,\varepsilon)}{\partial
\varepsilon_\alpha}\right \vert_{\varepsilon = 0}, \ \xi^i_\alpha =
\left. \frac{\partial X_i(t,x,\varepsilon)}{\partial
\varepsilon_\alpha}\right \vert_{\varepsilon = 0}, \ \eta^j_\alpha =
\left. \frac{\partial U_j(t,x,u,\varepsilon)}{\partial
\varepsilon_\alpha}\right \vert_{\varepsilon = 0}$.
%\end{equation}

In the extended space $\mathbb{R}^{n+m+q+1}$ of the variables
$(t,x,u,S)$ (hereafter $S = (S_1, ..., S_q)$ ), the Lie algebra
$L_N$ defines the  Lie group $\widetilde{G}_N$:
\begin{equation}\label{7}
t^{\ast} = T(t,x,\varepsilon), \ x_i^{\ast} = X_i(t,x,\varepsilon),
\ u^{\ast}_j = U_j(t,x,u,\varepsilon), \ S^{\ast}_b = S_b,
\end{equation}
where $i = 1, \ldots ,m, \ j = 1, \ldots, n, \ b = 1, \ldots, q$.

Now we propose a new definition, which is based on  the standard
definition of differential equation invariance as an invariant
manifold ${\cal{M}}$ \cite{olv} and generalizes the previous
definitions of BVP invariance (see, e.g., \cite{ben-olv-82, b-k, bl-anco02, ibr92}).

\begin{definition}  The BVP of the form (\ref{1})--(\ref{4}) is called invariant with respect to the  Lie group $\widetilde{G}_N$ (\ref{7}) if

\begin{itemize}
\item[(a)] the manifold determined by equation (\ref{1}) in the space of variables $\left(t,x,u, \ldots, u_x^{(k)}\right)$
 is invariant with respect to the
$k$th-order prolongation of the group $G_N$;
\item[(b)]each  manifold determined by conditions (\ref{2}) with
  any fixed number $a$  is invariant with respect
to the $k_a$th-order prolongation of the group $G_N$ in the space
of variables $\left(t,x,u, \ldots, u_x^{(k_a)}\right)$ , where $k_a
= \max \{k_{a}^j, \ j = 1, \ldots, n_a \}$;
\item[(c)] each  manifold determined by conditions (\ref{3}) with
 any fixed number $b$  is invariant with respect
to the $k_b$th-order prolongation of the group $\widetilde{G}_N$ in
the space of variables $\left(t,x,u, \ldots, u_x^{(k_b)}, S_b,
\ldots , S^{(k_b)}_b \right)$ , where $k_b=\max \{k_{b}^l, \ K_b^l,
\ l =1,\ldots,n_b \}$;
\item[(d)]each  manifold determined by conditions (\ref{4}) with
 any fixed number $c$  is invariant with respect
to the $k_c$th-order prolongation of the group $G_N$ in the space
of variables $\left(t,x,u, \ldots, u_x^{(k_c)}\right)$ ,
 where $k_c = \max \{k_{c}^m, \ m = 1, \ldots, n_c \}$.
\end{itemize}
\end{definition}

\begin{definition} The functions $u_j = \Phi_j(t,x), \,j = 1, \ldots,
n$ and $S_b = \Psi_b(t,x),\, b = 1, \ldots, q$  form  an invariant
solution $(u,S)$ of the BVP of the form (\ref{1})--(\ref{4}) corresponding to the
Lie group (\ref{7})~if
\begin{itemize}
\item[(i)] $(u,S)$ satisfies  equations  (\ref{1}) and conditions (\ref{1})--(\ref{4});
\item[(ii)] the manifold  ${\cal{M}} = \{u_j = \Phi_j(t,x), \ j = 1, \ldots, n; \,
S_b = \Psi_b(t,x), \ b = 1, \ldots, q \}$ is an invariant manifold
of the Lie group (\ref{7}).
\end{itemize}
\end{definition}

\begin{remark} Definition 1 can be straightforwardly generalized
on BVPs  with  governing systems of equations of hyperbolic,
elliptic  and mixed types. However, one should additionally assume
that $n$-component governing system of PDEs is presented in a
'canonical' form (some authors uses the natation 'involution form'
in this context), i.e. one possesses the simplest form and there are
no non-trivial differential consequences.
\end{remark}

\medskip

  If the system of differential equations   contain  arbitrary  functions as coefficients
 (formally speaking,  they  can be  constants),
the group classification problem arises.
Such problems  was formulated and
solved for a class of non-linear heat equations (NHEs) in a pioneering
work  \cite{ovs-59} (see also  \cite{ovs}).
At the present time,  there are      algorithms for rigorous solving   group classification problems (see, e.g., \cite{ch-se-ra-08} and references cited therein), which  were successfully
applied to different classes of PDEs.
Thus, if   system   (\ref{1})  and/or the
boundary conditions  (\ref{2})--(\ref{4}) contain arbitrary  functions as coefficients,
then we  should formulate  and solve  the
group classification problem for the BVP of the form (\ref{1})--(\ref{4}).

We propose the following  algorithm of the group classification for
the class of  BVPs (\ref{1})--(\ref{4}):
\begin{itemize}
\item[(I)] to construct  the equivalence group $E_{\mathrm{eq}}$ of local
transformations, which transform the governing system of equations
into itself;
\item[(II)] to extend the space of $E_{\mathrm{eq}}$ action on the variables $S =
(S_1, ..., S_q)$ by adding the identity transformations for  them
(see the last formula in (\ref{7})) and  to denote the group
obtained as $\widetilde{E}_{\mathrm{eq}}$;
\item[(III)] to find the equivalence group $\widetilde{E}_{\mathrm{eq}}^{\mathrm{BVP}}$
of local transformations, which transform the class of BVPs
(\ref{1})--(\ref{4}) into itself, one extends the space of  the
$\widetilde{E}_{\mathrm{eq}}$ action on the prolonged  space, where all
arbitrary elements arising in boundary conditions
(\ref{2})--(\ref{4}) are treated as new variables.
\item[(IV)] to perform the group classification of the
governing system (\ref{1}) up to local transformations generated by
the  group $\widetilde{E}_{\mathrm{eq}}^{\mathrm{BVP}}$;
\item[(V)] using Definition 1, to find the principal group of invariance
$\widetilde{G}^{0}$, which is admitted by each  BVP  belonging to
the class in question;
\item[(VI)] using Definition 1 and the results obtained at step (IV),
to  describe  all possible $\widetilde{E}_{\mathrm{eq}}^{\mathrm{BVP}}$-inequivalent
BVPs of the form (\ref{1})--(\ref{4}) admitting   maximal invariance
groups of higher dimensionality than $\widetilde{G}^{0}$.
\end{itemize}

\section{\bf Lie invariance of a class of (1+3)-dimensional nonlinear BVPs with free boundaries}

\subsection{\bf  Mathematical model of  melting and evaporation under a powerful flux of energy}

Consider the process of melting and evaporation in  half-space
$\Omega = \{\textbf{x} = (x_1, x_2, x_3): x_3
> 0\}$ occupied by a solid material,
 when its surface (initially it is the plane $x_3=0$) is exposed
to a powerful flux of energy.   We neglect the initial short-time
non-equilibrium stage of the process and consider the process at the
stage when three phases already take place and  assume that this
occurs at  any moment  $t \in \mathfrak{T}=(t_{\ast},+
\infty)$, where $t_{\ast}$ is a positive real number. Thus, the heat
transfer domain $\Omega(t)= \Omega \times \mathfrak{T}$ consists of
three sub-domains occupied by the gas, liquid and solid phases, which
 will be denoted by $\Omega_0(t)$, $\Omega_1(t)$  and
$\Omega_2(t),$ respectively, and the phase division boundary
surfaces, $S_1(t, \textbf{x})=0$ and  $S_2(t, \textbf{x})=0$ (see
Fig.1). In other words, the domain $\Omega(t)$ admits the disjoint
decomposition

\begin{figure}
\begin{center}
\vspace{1cm}
\includegraphics[width=8cm]{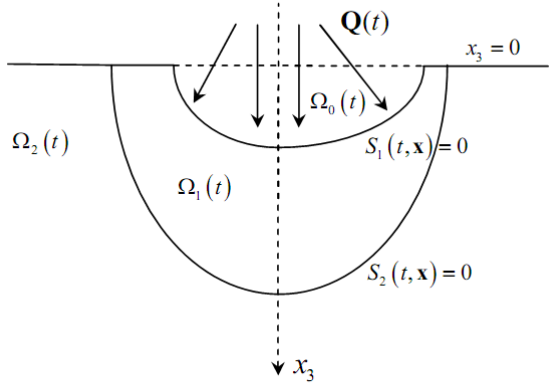}
\caption{A scheme for the process of melting and evaporation of a solid material which is exposed to a powerful energy flux.}
\end{center}
\end{figure}

\[
\Omega(t) = \Omega_0(t) \cup \Gamma_1(t) \cup \Omega_1(t) \cup
\Gamma_2(t) \cup \Omega_2(t),
\]
where
\[
\Gamma_k(t) = \{(t, \textbf{x}): S_k(t, \textbf{x}) = 0, \ t \in
\mathfrak{T}, \ \textbf{x} \in \Omega \}, \ k = 1,2,
\]
\[
\Omega_0(t) = \{(t, \textbf{x}): S_1(t, \textbf{x}) < 0, \ S_2(t,
\textbf{x}) < 0, \ t \in \mathfrak{T}, \ \textbf{x} \in \Omega \},
\]
\[
\Omega_1(t) = \{(t, \textbf{x}): S_1(t, \textbf{x}) > 0, \ S_2(t,
\textbf{x}) < 0, \ t \in \mathfrak{T}, \ \textbf{x} \in \Omega \},
\]
\[
\Omega_2(t) = \{(t, \textbf{x}): S_1(t, \textbf{x}) > 0, \ S_2(t,
\textbf{x}) > 0, \ t \in \mathfrak{T}, \ \textbf{x} \in \Omega \}.
\]

Let us consider a class of (1+3)-dimensional nonlinear BVPs of the
Stefan type
 used to describe melting and
evaporation of materials in the case when their surface is exposed
to a powerful flux of energy \cite{anisimov70, ready, gupta, ch-od91}:
\begin{eqnarray}
& & \nabla \left(\lambda_{1}(T_{1}) \nabla T_1 \right) =
C_{1}(T_{1}) \frac{\partial
T_{1}}{\partial t}, \ \ (t, \textbf{x}) \in \Omega_1(t), \label{8} \\
& & \nabla \left(\lambda_{2}(T_{2}) \nabla T_2 \right) =
C_{2}(T_{2}) \frac{\partial
T_{2}}{\partial t}, \ \ (t, \textbf{x}) \in \Omega_2(t), \label{9} \\
 & & \qquad S_{1}(t,\textbf{x}) = 0:\ \lambda_{1}(T_{v})
\frac{\partial T_{1}}{\partial \textbf{n}_1} = H_v \textbf{V}_1
\cdot \textbf{n}_1- \textbf{Q}(t) \cdot \textbf{n}_1, \ T_1 = T_v,\label{10} \\
& & \qquad S_{2}(t,\textbf{x}) = 0: \ \lambda_{2}(T_{m})
\frac{\partial T_{2}}{\partial \textbf{n}_2} = \lambda_{1}(T_{m})
\frac{\partial T_{1}}{\partial \textbf{n}_2} + H_m \textbf{V}_2
\cdot \textbf{n}_2,\ T_{1} = T_{2} = T_{m},\label{11}
\\ & & \qquad |\textbf{x}| = +\infty: \ T_{2} = T_{\infty}, \ \ t \in \mathfrak{T},\label{12}
\end{eqnarray}
where $T_v$, $T_{m}$ and $T_{\infty}$ are the known temperatures of
evaporation, melting and solid phases of the material,
respectively; $\lambda_{k}(T_k), \, k = 1,2$
 are the  positive thermal conductivities;
 $C_k(T_k)$, $H_v$, $H_m$ are the  positive specific heat
values per unit volume; $\textbf{Q}(t) = (Q_1(t),Q_2(t),Q_3(t))$ is
 the energy flux being absorbed by the material;
$S_{k}(t,\textbf{x}) = 0, \, k = 1,2$ are the phase division
boundary surfaces to be found; $\textbf{V}_k(t,\textbf{x}), \, k =
1,2$ are the phase division boundary velocities; $\textbf{n}_k, \, k
= 1,2$  are  the unit outward normals to the surfaces
$S_k(t,\textbf{x}) = 0, \, k = 1,2$; $T_{k}(t,\textbf{x}), \, k =
1,2$ are the unknown temperature fields; $\nabla =
\left(\frac{\partial}{\partial x_1}, \frac{\partial}{\partial x_2},
\frac{\partial}{\partial x_3} \right)$; the subscripts   $k = 1$ and
$k =2$ correspond to the liquid and solid phases, respectively.

Here equations (\ref{8}) and (\ref{9}) are basic and  describe the heat
transfer process in liquid and solid phases, the boundary conditions
(\ref{10}) present evaporation dynamics on the surface $S_{1} = 0$, and
the boundary conditions (\ref{11}) are the well-known Stefan
conditions on the surface $S_{2} = 0$ dividing the liquid and solid
phases. Since  the liquid  phase thickness is considerably less than
the solid phase thickness, one may use the Dirichlet condition
(\ref{12}), where $T_{\infty}$  can be treated as  the initial
temperature of material.  We  also assume that  the gas phase does
not interact with  liquid and solid phases; hence  the problem in
question does not involve any equation for the gas phase.

From the mathematical and  physical points of view,  we should
impose the same additional conditions on the functions and constants
arising in the BVP class in question, which guarantee existing
classical solutions. Namely,  we assume that  all functions in
(\ref{8})--(\ref{12})  are sufficiently smooth; the free boundary
surfaces $S_k(t, \textbf{x}) = 0$   satisfy the restrictions
$\frac{\partial S_k}{\partial t} \neq 0$ and $|\nabla S_k| \neq 0$,
$k = 1,2$; the projection of the heat flux vector $\textbf{Q}(t)$ on
the normal $\textbf{n}_1$  is nonzero, i.e. $\textbf{Q}(t) \cdot
\textbf{n}_1 \neq 0$,   and
$\textbf{V}_k \cdot \textbf{n}_k \neq~0, \  k=1,2 $.
 Finally, the constants $T_v$, $ T_m$ and $T_{\infty}$  satisfy
the natural  inequalities $T_v > T_m >T_{\infty}$.

First of all, we simplify the governing system of equations
(\ref{8}) and (\ref{9}) using the Goodman substitution \cite{kozdoba, goodman}
\begin{equation}\label{13}
u = \phi_1(T_1) \equiv \int\limits_0^{T_{1}} {C_{1}(\zeta)}\,d\zeta,
\quad v = \phi_2(T_2) \equiv \int\limits_0^{T_{2}}
{C_{2}(\xi)}\,d\xi.
\end{equation}
Substituting (\ref{13}) into (\ref{8})--(\ref{12}) and making the
relevant calculations, we arrive at the equivalent class of BVPs
\begin{eqnarray}
& & \frac{\partial u}{\partial t} = \nabla
\left(d_{1}(u) \nabla u \right), \ \ (t, \textbf{x}) \in \Omega_1(t),\label{14} \\
& & \frac{\partial v}{\partial t} = \nabla \left(d_{2}(v)
\nabla v \right) , \ \ (t, \textbf{x}) \in \Omega_2(t),\label{15} \\
 & & \qquad S_{1}(t,\textbf{x}) = 0:\ d_{1v}
\frac{\partial u}{\partial \textbf{n}_1} = H_v \textbf{V}_1
\cdot \textbf{n}_1- \textbf{Q}(t) \cdot \textbf{n}_1, \ u = u_v,\label{16} \\
& & \qquad S_{2}(t,\textbf{x}) = 0: \ d_{2m} \frac{\partial
v}{\partial \textbf{n}_2} = d_{1m} \frac{\partial u}{\partial
\textbf{n}_2} + H_m \textbf{V}_2 \cdot \textbf{n}_2,\ u = u_m, \ v =
v_m,\label{17}
\\ & & \qquad |\textbf{x}| = +\infty: \ v = v_{\infty}, \ \ t \in \mathfrak{T},\label{18}
\end{eqnarray}
where $u_v = \int\limits_0^{T_{v}} {C_{1}(\zeta)}\,d\zeta$, $u_m =
\int\limits_0^{T_{m}} {C_{1}(\zeta)}\,d\zeta$, $v_m =
\int\limits_0^{T_{m}} {C_{2}(\xi)}\,d\xi$, $v_{\infty} =
\int\limits_0^{T_{\infty}} {C_{2}(\xi)}\,d\xi$; $d_1(u) =
\frac{\lambda_1(\phi^{-1}_1(u))}{C_1(\phi^{-1}_1(u))}$, $d_2(v) =
\frac{\lambda_2(\phi^{-1}_2(v))}{C_2(\phi^{-1}_2(v))}$; $d_{1v} =
d_1(u_v)$, $d_{1m} = d_1(u_m)$, $d_{2m} = d_2(v_m)$ (here
$\phi^{-1}_k, \, k=1,2$ are the  inverse functions of $\phi_k$, the
functions $d_1(u)$ and $d_2(v)$ are strictly positive and $u_v \neq
u_m$, $v_m \neq v_{\infty}$).

\subsection{\bf  Group classification of the basic equations  (\ref{14})--(\ref{15}) }

One  sees that the BVP of the form (\ref{14})--(\ref{18})  consists of   two
standard NHEs  with  arbitrary smooth
functions $d_1(u)$ and $d_2(v)$  and  the boundary conditions
(\ref{16})--(\ref{18}) containing   an arbitrary vector function
$\textbf{Q}(t)$ and  a number of  arbitrary  parameters. Thus, we
deal with a class of BVPs, and to  carry out the group classification
  the algorithm formulated in Section 2  can be used.

According to item (I), we need to   find the group of equivalent
transformations of the   non-coupled system  of  NHEs
(\ref{14})--(\ref{15}). Note that this group is well-known in the
case of  a single NHE  (see, e.g., \cite{dor-svi}). However, one
cannot   extend this result in  the case of system
(\ref{14})--(\ref{15})  in a formal way because  the group obtained
may be incomplete. Thus, we carefully check the group of
equivalent transformations for the non-coupled system
(\ref{14})--(\ref{15}).

\begin{lema}
The  equivalence transformations  group $E_{\mathrm{eq}}$  of  system
(\ref{14})--(\ref{15}) consists of the  group $\mathcal{E}_{\mathrm{eq}}$ of
continuous equivalence transformations
\begin{eqnarray}
& & \nonumber \bar{t} = \alpha t + \gamma_0, \ \bar{X} = \beta A_i(\beta_1) A_j(\beta_2) A_k(\beta_3) X + \Gamma \ (i,j,k = 1,2,3; \ i \neq j, i \neq k, j \neq k), \\
& & \nonumber \bar{u} = \delta_1 u + \gamma_4, \ \bar{v} = \delta_2
v + \gamma_5, \ \bar{d}_1 = \frac{\beta^2}{\alpha} \ d_1, \
\bar{d}_2 = \frac{\beta^2}{\alpha} \ d_2,
\end{eqnarray}
and  the group of  discrete equivalence transformations
\begin{itemize}
\item[1)] $t \rightarrow -t$, \ $x_i \rightarrow  (-1)^jx_i \ (i = 1, 2, 3, \, j=0,1)$, \ $u \rightarrow -u$, \ $v \rightarrow -v$, \ $d_1 \rightarrow
-d_1$, \ $d_2 \rightarrow -d_2$;
\item[2)]  $t \rightarrow t$, \ $x_i \rightarrow  (-1)^jx_i \ (i = 1, 2, 3, \, j=0,1)$, \ $u \rightarrow v$, \ $v \rightarrow u$, \ $d_1 \rightarrow
d_2$, \ $d_2 \rightarrow d_1$.
\end{itemize}
Here $\alpha>0, \beta>0, \beta_1, \ldots, \beta_3, \gamma_0, \ldots,
\gamma_5, \delta_1>0, \delta_2>0$ are arbitrary  constants;
\[
A_1(\theta) = \left (\begin{array}{ccc} \cos\theta & \sin\theta & 0
\\ - \sin\theta & \cos\theta & 0 \\ 0 & 0 & 1 \end{array} \right ),
 A_2(\theta) = \left (\begin{array}{ccc} \cos\theta & 0 &
\sin\theta \\ 0 & 1 & 0 \\ -\sin\theta & 0 & \cos\theta \end{array}
\right),  A_3(\theta) = \left (\begin{array}{ccc} 1 & 0 & 0 \\ 0 &
\cos\theta & \sin\theta \\ 0 & - \sin\theta & \cos\theta \end{array}
\right)
\]
are the matrix of rotation in space, and
\[
X = \left (\begin{array}{c} x_1 \\ x_2 \\ x_3 \end{array} \right), \
\Gamma = \left (\begin{array}{c} \gamma_1 \\ \gamma_2 \\ \gamma_3
\end{array} \right).
\]
\end{lema}

{\bf Proof.} To  find  the group $\mathcal{E}_{\mathrm{eq}}$, we  use  the
standard infinitesimal method \cite{ovs, ibr91}, i.e., search for
the generators of  the form
\[
Y = \xi^0 \frac{\partial}{\partial t}+\xi^a \frac{\partial}{\partial
x_a} + \eta^1 \frac{\partial}{\partial u} + \eta^2
\frac{\partial}{\partial v} + \mu^1 \frac{\partial}{\partial d_1} +
\mu^2 \frac{\partial}{\partial d_2},
\]
where $a = 1, 2, 3$ (hereafter, summation is assumed from 1 to 3
over repeated indices $a$ or $b$). \noindent The generator $Y$
defines the group $\mathcal{E}_{\mathrm{eq}}$ of equivalence transformations
\[
\bar{t} = \phi(t, \textbf{x}, u, v), \ \bar{x}_a = \psi_a(t,
\textbf{x}, u, v), \ \bar{u} = \Phi_1(t, \textbf{x}, u, v), \
\bar{v} = \Phi_2(t, \textbf{x}, u, v),
\]
\[
\bar{d_1} = \Psi_1(t, \textbf{x}, u, v, d_1, d_2), \ \bar{d_2} =
\Psi_2(t, \textbf{x}, u, v, d_1, d_2)
\]
for the class of systems (\ref{14})--(\ref{15}) iff $Y$
obeys the condition of invariance of the following system:
\begin{equation}\label{lem1}
\frac{\partial u}{\partial t} - \nabla \left(d_{1} \nabla u \right)
= 0, \ \frac{\partial d_1}{\partial t} = 0, \ \frac{\partial
d_1}{\partial x_a} = 0, \ \frac{\partial d_1}{\partial v} = 0,
\end{equation}
\begin{equation}\label{lem2}
\frac{\partial v}{\partial t} - \nabla \left(d_{2} \nabla v \right)
= 0, \ \frac{\partial d_2}{\partial t} = 0, \ \frac{\partial
d_2}{\partial x_a} = 0, \ \frac{\partial d_2}{\partial u} = 0.
\end{equation}
Here, the  variables  $u$ and $v$  are considered in the space $(t,
\textbf{x})$, while  $d_k$  in the extended space $(t, \textbf{x},
u, v)$. The coordinates $\xi^0, \xi^a$ and $\eta^k$ of the operator $Y$
are the functions of the variables  $t, \textbf{x}, u, v$, while the
coordinates $\mu^k$ are the functions of $t, \textbf{x}, u, v, d_1,
d_2$. Thus, the  invariance criterium
 for  system (\ref{lem1})--(\ref{lem2}) is given by the formulae
\begin{equation}\label{lem3}
Y^{(2)} \left.\left(\frac{\partial u}{\partial t} - \nabla
\left(d_{1} \nabla u \right)\right)\right\vert_{\cal S} = 0, \
Y^{(2)} \left.\left(\frac{\partial d_1}{\partial
t}\right)\right\vert_{\cal S} = 0, \ Y^{(2)}
\left.\left(\frac{\partial d_1}{\partial
x_a}\right)\right\vert_{\cal S} = 0, \ Y^{(2)}
\left.\left(\frac{\partial d_1}{\partial v}\right)\right\vert_{\cal
S} = 0,
\end{equation}
\begin{equation}\label{lem4}
Y^{(2)} \left.\left(\frac{\partial v}{\partial t} - \nabla
\left(d_{2} \nabla v \right)\right)\right\vert_{\cal S} = 0, \
Y^{(2)} \left.\left(\frac{\partial d_2}{\partial
t}\right)\right\vert_{\cal S} = 0, \ Y^{(2)}
\left.\left(\frac{\partial d_2}{\partial
x_a}\right)\right\vert_{\cal S} = 0, \ Y^{(2)}
\left.\left(\frac{\partial d_2}{\partial v}\right)\right\vert_{\cal
S} = 0,
\end{equation}
where $S$ is the manifold, defined by system
(\ref{lem1})--(\ref{lem2}), $Y^{(2)}$ is the second prolongation of
the operator $Y$ calculated via the known formulae \cite{ovs,
ibr91}. Using these formulae and  (\ref{lem3})--(\ref{lem4}) and
carrying out  the relevant calculations  one obtains  the
13-dimensional Lie algebra
 with the basic operators
\[
Y_1 = \frac{\partial}{\partial t}, \ Y_2 = \frac{\partial}{\partial
x_1}, \ Y_3 = \frac{\partial}{\partial x_2}, \ Y_4 =
\frac{\partial}{\partial x_3},
\]
\[
Y_5 = x_2 \frac{\partial}{\partial x_1} - x_1
\frac{\partial}{\partial x_2}, \ Y_6 = x_3 \frac{\partial}{\partial
x_1} - x_1 \frac{\partial}{\partial x_3}, \ Y_7 = x_3
\frac{\partial}{\partial x_2} - x_2 \frac{\partial}{\partial x_3},
\]
\[
Y_8 = \frac{\partial}{\partial u}, \ Y_9 = \frac{\partial}{\partial
v}, \ Y_{10} = u \frac{\partial}{\partial u}, \ Y_{11} = v
\frac{\partial}{\partial v},
\]
\[
Y_{12} = t \frac{\partial}{\partial t} - d_1
\frac{\partial}{\partial d_1} - d_2 \frac{\partial}{\partial d_2}, \
Y_{13} = x_a \frac{\partial}{\partial x_a} + 2 d_1
\frac{\partial}{\partial d_1} + 2 d_2 \frac{\partial}{\partial d_2}.
\]
One easily checks that  this algebra  generates  the  group
$\mathcal{E}_{\mathrm{eq}}$.

Finally, to obtain the   group of equivalence transformations
$E_{\mathrm{eq}}$, we should add the  discrete transformations 1) and 2)
listed in Lemma 1.
 It is easy to verify by the direct calculations   that system (\ref{14})--(\ref{15}) is invariant under
 these discrete transformations.

The proof is now complete. $\blacksquare$

According to item (II)  of the algorithm, we should now construct
the group $\widetilde{E}_{\mathrm{eq}}$ by adding the identity
transformations $\widetilde{S}_1=S_1, \widetilde{S}_2=S_2$. To
realize item (III), one needs to  apply  the transformations
generated by the group
 $\widetilde{E}_{\mathrm{eq}}$ to  the boundary conditions
(\ref{16})--(\ref{18})  and to find the group
 $\widetilde{E}^{\mathrm{BVP}}_{\mathrm{eq}}$.
 The result  obtained  is   formulated  as  follows.

\begin{lema}
The   class of BVPs (\ref{14})--(\ref{18}) admits the group of
equivalence  transformations $\widetilde{E}^{\mathrm{BVP}}_{\mathrm{eq}}$:
\begin{eqnarray}
& & \nonumber \tilde{t} = \alpha t + \gamma_0, \ \widetilde{X} = \beta A_i(\beta_1) A_j(\beta_2) A_k(\beta_3) X + \Gamma \ (i,j,k = 1,2,3; \ i \neq j, i \neq k, j \neq k),\\
& & \nonumber \tilde{u} = \delta_1 u + \gamma_4, \ \tilde{v} =
\delta_2 v + \gamma_5, \ \widetilde{S}_1 = S_1, \ \widetilde{S}_2 =
S_2, \ \tilde{d}_1 = \frac{\beta^2}{\alpha} \ d_1,
\ \tilde{d}_2 = \frac{\beta^2}{\alpha} \ d_2, \\
& & \nonumber \tilde{d}_{1v} = \frac{\beta}{\delta_1} \ d_{1v}, \
\tilde{d}_{1m} = \frac{\beta}{\delta_1} \ d_{1m}, \ \tilde{d}_{2m}
= \frac{\beta}{\delta_2} \ d_{2m}, \ \widetilde{H}_v = \frac{\alpha}{\beta} H_v,
\ \widetilde{H}_m = \frac{\alpha}{\beta} H_m, \\
& & \nonumber \tilde{u}_v = \delta_1 u_v + \gamma_4, \ \tilde{u}_m =
\delta_1 u_m + \gamma_4, \ \tilde{v}_m = \delta_2 v_m + \gamma_5, \ \tilde{v}_{\infty} = \delta_2 v_{\infty} + \gamma_5\\
& & \nonumber \widetilde{Q} = A_i(\beta_1) A_j(\beta_2)
A_k(\beta_3) Q \ (i,j,k = 1,2,3; \ i \neq j, i \neq k, j \neq k),
\end{eqnarray}
with arbitrary coefficients $\alpha, \beta, \beta_1, \ldots,
\beta_3, \gamma_0, \ldots, \gamma_5, \delta_1, \delta_2$ obeying
only the condition
\[
\alpha \beta \delta_1 \delta_2 \neq 0.
\]
\end{lema}

\begin{table}
\caption {Lie algebras of the NHE system (\ref{14})--(\ref{15})
($k_1,k_2, m$ and $n$ are arbitrary non-zero constants, $b < a = 1,
2, 3$; while $\alpha(t,x)$ and $\beta(t,x)$ are arbitrary solutions
of the linear heat equations $\alpha_t = k_1 \alpha_{x x}$ and
$\beta_t = k_2 \beta_{x x}$, respectively.)}\label{tab1}
\renewcommand{\arraystretch}{1.5}
\tabcolsep=5 pt
\begin{center}
\begin{tabular}{cccl}
  \hline\hline
  {no} & {$d_1(u)$} & {$d_2(v)$} & {Basic operators of MAI}\\
  \hline
  1. & $\forall$ & $\forall$ & $ AE(1,3) = \langle \partial_{t}, \partial_{x_a}, x_a \partial_{x_b} - x_b \partial_{x_a}, 2t \partial_{t} + x_a \partial_{x_a} \rangle$\\
  2. & $k_1$ & $\forall$ & $ AE(1,3), u \partial_u, \alpha(t,\textbf{x}) \partial_u $\\
  3. & $\forall$ & $k_2$ & $ AE(1,3), v \partial_v, \beta(t,\textbf{x}) \partial_v $\\
  4. & $e^u$ & $e^v$ & $ AE(1,3), x_a \partial_{x_a} + 2 \partial_{u} +  2 \partial_{v}$  \\
  5. & $e^u$ & $v^m$& $ AE(1,3), x_a \partial_{x_a} + 2 \partial_u + \frac{2}{m} v \partial_{v} $ \\
  6. & $u^n$ & $e^v$& $ AE(1,3), x_a \partial_{x_a} + \frac{2}{n} u \partial_u + 2 \partial_{v} $ \\
  7. & $u^n$ & $v^m$& $ AE(1,3),  D=x_a \partial_{x_a} + \frac{2}{n} u \partial_u + \frac{2}{m} v \partial_{v} $ \\
  8. & $u^{-\frac{4}{5}}$ & $v^{-\frac{4}{5}}$ & $ AE(1,3), |\textbf{x}|^2 \partial_{x_b} - 2 x_b x_a \partial_{x_a} + 5 x_b u \partial _u + 5 x_b v \partial_v, \,
  %x_a \partial_{x_a} - \frac{5}{2} u \partial_u - \frac{5}{2} v   \partial_{v},
   D$ with $m=n= -\frac{4}{5} $\\
  9. & $k_1$ & $k_2$ & $AE(1,3), u \partial_u, v \partial_v, \alpha(t,\textbf{x}) \partial_u, \beta(t,\textbf{x}) \partial_v, G_a=t \partial_{x_a} - x_a \left(\frac{1}{2k_1} u \partial_u + \frac{1}{2k_2}v \partial_v \right)
  $,\\ &  &  &  $\Pi= t^2 \partial_t +  t x_a \partial_{x_a} - \frac{1}{4k_1}(|\textbf{x}|^2 + 6 k_1 t) u \partial_u - \frac{1}{4k_2}(|\textbf{x}|^2 + 6 k_2 t) v \partial_v$ \\
   10. & $k_1$ & $k_1$ & $AE(1,3), u \partial_u, v \partial_v, v \partial_u, u \partial_v, \alpha(t,\textbf{x}) \partial_u, \beta(t,\textbf{x}) \partial_v, G_a$  and  $\Pi$  with $ k_2=k_1$ \\
  \hline\hline
\end{tabular}
\end{center}
\end{table}

According to item (IV)  of the algorithm, we should now perform the
group classification of the governing system (\ref{14})--(\ref{15}) up to local transformations generated by the  group $\widetilde{E}_{\mathrm{eq}}^{\mathrm{BVP}}$.
The result can be presented as follows.

\begin{theorem}
All possible maximal algebras of invariance (MAIs) (up to equivalent
representations generated by transformations from the group
$\widetilde{E}^{\mathrm{BVP}}_{\mathrm{eq}}$) of system (\ref{14})--(\ref{15}) for any
fixed vector function  $(d_1, d_2)$ with strictly positive functions $d_1(u)$
and $d_2(v)$ are presented in Table~\ref{tab1}. Any other system of
the form (\ref{14})--(\ref{15}) is reduced to one of those with
diffusivities from Table~\ref{tab1} by an equivalence transformation
from the group $\widetilde{E}^{\mathrm{BVP}}_{\mathrm{eq}}$.
\end{theorem}

{\bf Proof.} First of all, we  remind the reader that
 a complete description of Lie symmetries  for the class of
multi-dimensional nonlinear reaction-diffusion systems
\begin{eqnarray}
& & \nonumber \frac{\partial u}{\partial t} = \nabla
\left(d_{1}(u) \nabla u \right) + F_1(u,v), \\
& & \nonumber \frac{\partial v}{\partial t} = \nabla \left(d_{2}(v)
\nabla v \right) + F_2(u,v),
\end{eqnarray}
where $F_1$ and $F_2$ are the arbitrary nonzero smooth functions, was
obtained in \cite{ch-king00, ch-king03} (the case of
constant diffusivities) and \cite{ch-king06}  (the case of
non-constant diffusivities). Nevertheless, the detailed examination
of the special   case $F_1 =F_2 = 0$  was omitted in the papers
\cite{ch-king00, ch-king03, ch-king06}, we can use the  relevant
system of the determining equations with $F_1 =F_2 = 0$ to find  all
possible MAIs of system
(\ref{14})--(\ref{15}). Let us assume that the MAI   in question is
generated by the infinitesimal operator
\[
X = \xi^0(t, \textbf{x}, u, v) \frac{\partial}{\partial t}+\xi^a(t,
\textbf{x}, u, v) \frac{\partial}{\partial x_a} + \eta^1(t,
\textbf{x}, u, v) \frac{\partial}{\partial u} + \eta^2(t,
\textbf{x}, u, v) \frac{\partial}{\partial v},
\]
where  $ \xi^0, \  \xi^a, \  \eta^1$ and $ \eta^2$   are  the
unknown smooth   functions.  The form of the system of determining
equations to find these functions essentially depends on the
derivatives $\frac{d d_1}{d u}$ and $\frac{d d_2}{d v}$. Let us
consider the most general case $\frac{d d_1}{d u} \cdot
\frac{d d_2}{d v} \neq 0$. The  relevant  system of determining
equations has the form (see  formulae (11)--(16) with $F_1 =F_ 2 =
0$ in  \cite{ch-king06})
\begin{equation}\label{p6}
\xi^0_{x_a} = \xi^0_{u} = \xi^0_{v} = 0, \ \xi^a_{u} = \xi^a_{v} =
\eta^1_v = \eta^2_u = 0, \ a = 1, 2, 3,
\end{equation}
\begin{equation}\label{p7}
\xi^1_{x_1} = \xi^2_{x_2} = \xi^3_{x_3}, \ \xi^a_{x_b} + \xi^b_{x_a}
= 0, \ b < a = 1, 2, 3,
\end{equation}
\begin{equation}\label{p8}
\xi^0_t = 2 \xi^a_{x_a} - \eta^1 \frac{d}{d u} \log d_1, \ \xi^0_t =
2 \xi^a_{x_a} - \eta^2 \frac{d}{d v} \log d_2, \ a = 1, 2, 3,
\end{equation}
\begin{equation}\label{p9}
2 \frac{\partial^2 \eta^1}{\partial x_a \partial u} + 2 \eta^1_{x_a}
\frac{d}{d u} \log d_1 = \Delta \xi^a - \xi^a_t \left(d_1
\right)^{-1}, \ 2 \frac{\partial^2 \eta^2}{\partial x_a \partial v}
+ 2 \eta^2_{x_a} \frac{d}{d v} \log d_2 = \Delta \xi^a - \xi^a_t
\left(d_2 \right)^{-1},
\end{equation}
\begin{equation}\label{p10}
2 \xi^a_{x_a} - \xi^0_t - \eta^1_u = \eta^1_{u u} \left(\frac{d}{d
u} \log d_1 \right)^{-1} + \eta^1 \frac{d}{d u} \log \frac{d d_1}{d
u}, \ a = 1, 2, 3,
\end{equation}
\begin{equation}\label{p11}
2 \xi^a_{x_a} - \xi^0_t - \eta^2_v = \eta^2_{v v} \left(\frac{d}{d
v} \log d_2 \right)^{-1} + \eta^2 \frac{d}{d v} \log \frac{d d_2}{d
v}, \ a = 1, 2, 3,
\end{equation}
\begin{equation}\label{p12}
\eta^1_t - \Delta \eta^1 d_1 = 0, \ \eta^2_t - \Delta \eta^2 d_2 =
0.
\end{equation}

If  $d_1(u)$ and $d_2(v)$  are  the arbitrary smooth  functions, then
system (\ref{p6})--(\ref{p12})  can be easily solved  resulting the
eight-dimensional Lie  algebra $AE(1,3)$ with the  basic operators
\[
P_t = \frac{\partial}{\partial t}, \ P_a = \frac{\partial}{\partial
x_a}, \ J_{a b} = x_a P_b - x_b P_a, \ D_0 = 2 t
\frac{\partial}{\partial t} + x_a \frac{\partial}{\partial x_a}, \,
b < a = 1, 2, 3.
\]

Now, we should  find all possible pairs of the  function $d_1(u)$
and $d_2(v)$ leading to   extensions  of the  algebra $AE(1,3)$. It
is evident that equations (\ref{p6}) and (\ref{p8}) can be easily
solved and  the functions
\begin{equation}\label{p13}
\xi^0 = A(t) , \ \xi^a = B^a(t,x), \ \ a = 1, 2, 3,
\end{equation}
\begin{equation}\label{p14}
\eta^1 = (2 B^a_a - A_t) \left(\frac{d}{d u} \log d_1 \right)^{-1},
\ \eta^2 = (2 B^a_a - A_t) \left(\frac{d}{d v} \log d_2
\right)^{-1}, \ \ a = 1, 2, 3
\end{equation}
are obtained (here $A$ and $B^a$ are the arbitrary functions at  the
moment). However, the functions $B^a$ can be specified using
(\ref{p7}) as follows:
\begin{equation}\label{p15}
B^a = c_{0a} + c_{ab} x_b + k_a(t) |\textbf{x}|^2 - 2 k_b(t) x_b
x_a, \ a, b = 1, 2, 3,
\end{equation}
where $c_{ab} + c_{ba} = 0 \, (a \neq b), c_{11} = \ldots = c_{33}$ and
$c_{0a}, c_{ab}, k_a(t), k_b(t)$ are arbitrary functions.

Substituting formulae (\ref{p13}) and (\ref{p14}) in
(\ref{p10}) and (\ref{p11}) we arrive at the system of classification
equations to find pairs of  the functions $(d_1, d_2)$:
\begin{equation}\label{p16}
\frac{d^2}{du^2} \left(\frac{d}{d u} \log d_1 \right)^{-1} = 0, \
\frac{d^2}{dv^2} \left(\frac{d}{d v} \log d_2 \right)^{-1} = 0,
\end{equation}
otherwise, the conditions
\begin{equation}\label{p17}
2 B^a_a - A_t = 0, \ a = 1, 2, 3,
\end{equation}
must hold. However, substituting (\ref{p15}) in conditions
(\ref{p17}) and taking into account (\ref{p9}) and
(\ref{p13})--(\ref{p14}) we immediately obtain only the Lie
algebra $AE(1,3)$. System (\ref{p16}), which consists of two
independent ODEs, can be easy solved:
\begin{equation}\label{p18}
d_1(u) = \left \{\begin{array}{lll} D_1 (u + C_1)^{\alpha_1}, \\
D_1 \exp(\alpha_1 u), \end{array} \right.
\end{equation}
\begin{equation}\label{p19}
d_2(v) = \left \{\begin{array}{lll} D_2 (v + C_2)^{\alpha_2}, \\
D_2 \exp(\alpha_1 v), \end{array} \right.
\end{equation}
where $D_k \neq 0, \ \alpha_k \neq 0$ and $C_k$ are arbitrary
constants ($k = 1, 2$).

Substitutions the functions $\xi^0, \xi^a, \eta^1$ and $\eta^2$ from
(\ref{p13})--(\ref{p14}) in (\ref{p9}) we arrive at the equations
\[
B_t^a = - \Delta B^a \left (5 + 4 \frac{d^2}{d u^2} \log d_1 \right)
\frac{d d_1}{d u}, \ \ B_t^a = - \Delta B^a \left (5 + 4
\frac{d^2}{d v^2} \log d_2 \right) \frac{d d_2}{d v}.
\]
Since the functions $B_a, \ (a = 1, 2, 3)$ depend only on $t$ and
$x$, there are only two possibilities. The first one
is  $\Delta B^a
= 0$ and then, applying (\ref{p15}), we obtain
\[
B^a = c_{a b} x_b + c_{a 0}, \ \ c_{a b}, c_{a 0} \in \mathbb{R}.
\]
The second possibility, having $\Delta B^a \neq 0$, requires
\begin{equation}\label{p20}
\left (5 + 4 \frac{d^2}{d u^2} \log d_1 \right) \frac{d d_1}{d u} =
m_1, \ \ \left (5 + 4 \frac{d^2}{d v^2} \log d_2 \right) \frac{d
d_2}{d v} = m_2,
\end{equation}
where $m_1$ and $m_2$ are some constants. Using
(\ref{p18})--(\ref{p19}) it is easily seen  that conditions
(\ref{p20}) can be satisfied only for $m_1 = m_2 = 0$, and then
\begin{equation}\label{p21}
d_1(u) = D_1 \left(u + C_1 \right)^{- \frac{4}{5}}, \ \ d_1(v) = D_2
\left(v + C_2 \right)^{- \frac{4}{5}}.
\end{equation}
Thus,  we have obtained all possible forms of the functions $d_1(u)$ and
$d_2(v)$, namely formulae (\ref{p18})--(\ref{p19}) and (\ref{p21}),
leading to   extensions  of the  invariance  algebra $AE(1,3)$  of the nonlinear
system (\ref{14})--(\ref{15}), when both diffusivities are
non-constant.

Finally, taking into account the point transformations
from the group $\widetilde{E}^{\mathrm{BVP}}_{\mathrm{eq}}$, we immediately obtain
cases 4--8 from Table \ref{tab1}. Other possibility, when at least
one of the functions $d_1(u)$ and $d_2(v)$ is constant,  was
examined  in a similar way, and cases 2,3 and 9, 10 from Table
\ref{tab1} were  obtained.

The proof is now complete. $\blacksquare$

\medskip

\begin{remark} Cases 2 and 5 from Table \ref{tab1} are
equivalent  to 3 and 6, respectively, if one takes into
account the discrete transformations 2) from the group $E_{\mathrm{eq}}$.
However, these transformations do not belong to the
$\widetilde{E}^{\mathrm{BVP}}_{\mathrm{eq}}$, because the  boundary  conditions
(\ref{16}) and (\ref{18}) are not invariant under them.
\end{remark}

\subsection{\bf  Group classification of the class of  BVPs   (\ref{14})--(\ref{18}) }

We  note that   each MAI   from  Table \ref{tab1}
generates the corresponding
 maximal groups of invariance (MGI) with the transformations, which can be easily derived from the
basic  generators  listed therein. These transformations are
well-known  and   used below.

Taking this into account and  according to items  (V)  and (VI)    of the algorithm,
 we  formulate the  main  theorem  giving  the complete
list of  Lie  symmetries  of     the  BVP  class
(\ref{14})--(\ref{18}).

\begin{theorem}
The BVP of the form (\ref{14})--(\ref{18}), with  the arbitrary  given functions
$d_1(u)$, $d_2(v)$ ($d_1(u)\not= d_2(v)$) and $Q_a(t), \ a = 1, 2 , 3$, is invariant
 under the three-parameter Lie group (trivial Lie group) presented in
case 1 of Table \ref{tab2}. The MGI of any BVP of the form
(\ref{14})--(\ref{18}) does not  depend on the form of   $d_1(u)$ and
$d_2(v)$.  There are  only five BVPs  from  class
(\ref{14})--(\ref{18})  with  the  correctly specified  functions
$Q_a(t), \ a = 1, 2 , 3$ admitting the MGI of a higher dimensionality,
namely:  four- or five-parameter  Lie groups of  invariance  (up to
equivalent representations generated by equivalence transformations
from the group $\widetilde{E}^{\mathrm{BVP}}_{\mathrm{eq}}$).
These MGIs  and the relevant   functions
$Q_a(t), \ a = 1, 2 , 3$ are  presented in cases 2--6 of Table \ref{tab2}.
\end{theorem}

\begin{table}
\caption {Lie invariance of BVP (\ref{14})--(\ref{18})}
\label{tab2}
\renewcommand{\arraystretch}{1.5}
\tabcolsep=10pt
\begin{center}
\begin{tabular}{ccccl}
  \hline\hline
  {no} & {$Q_1(t)$} & {$Q_2(t)$} & {$Q_3(t)$} & {MGI}\\
  \hline
  1. & $ \forall $ & $ \forall $ & $ \forall $ & $ \widetilde{T}_1, \widetilde{T}_2, \widetilde{T}_3 $\\
  2. & $0$ & $0$ & $q(t)$ & $ \widetilde{T}_1, \widetilde{T}_2, \widetilde{T}_3, \widetilde{T}_5 $\\
  3. & $\Theta_1(\lambda t)$ & $\Theta_2(\lambda t)$ & $q_3$ & $ \widetilde{T}_1, \widetilde{T}_2, \widetilde{T}_3, \widetilde{T}_{6}$\\
  4. & $\frac{1}{\sqrt t} \ \Theta_1\left(\frac{1}{2} \lambda \log t\right)$ & $\frac{1}{\sqrt t} \ \Theta_2\left(\frac{1}{2} \lambda \log t\right)$ & $\frac{q_3}{\sqrt t}$ & $ \widetilde{T}_1, \widetilde{T}_2, \widetilde{T}_3, \widetilde{T}_{7}$\\
  5. & $0$ & $0$ & $q$ & $ \widetilde{T}_0, \widetilde{T}_1, \widetilde{T}_2, \widetilde{T}_{3}, \widetilde{T}_{5}$\\
  6. & $0$ & $0$ & $\frac{q}{\sqrt t}$ & $ \widetilde{T}_1, \widetilde{T}_2, \widetilde{T}_3, \widetilde{T}_{4}, \widetilde{T}_{5}$\\
  \hline\hline
\end{tabular}
\end{center}
\end{table}

\begin{remark} In Table \ref{tab2}, the  following  designations are
used: $q \neq 0$, $q_3$, $\lambda$ are arbitrary constants, $q(t) \neq 0$ is an arbitrary function while the functions
\[
\Theta_1(\tau) = q_1 \cos{\tau} + q_2 \sin{\tau}, \ \ \Theta_2(\tau) = - q_1 \sin{\tau} + q_2 \cos{\tau},
\]
where $q_1$, $q_2$ are arbitrary constants satisfying the condition $q_1^2 + q_2^2 \neq 0$ if $\tau \neq 0$.
\end{remark}

The explicit form of the transformations generating the MGI are:
\[
\widetilde{T}_0: t^{\ast} = t + \varepsilon_0, \ x_1^{\ast} = x_1, \
x_2^{\ast} = x_2, \ x_3^{\ast} = x_3, \ u^{\ast} = u, \ v^{\ast} =
v, \ S_1^{\ast} = S_1, \ S_2^{\ast} = S_2,
\]
\[
\widetilde{T}_1: t^{\ast} = t, \ x_1^{\ast} = x_1 + \varepsilon_1, \
x_2^{\ast} = x_2, \ x_3^{\ast} = x_3, \ u^{\ast} = u, \ v^{\ast} =
v, \ S_1^{\ast} = S_1, \ S_2^{\ast} = S_2,
\]
\[
\widetilde{T}_2: t^{\ast} = t, \ x_1^{\ast} = x_1, \ x_2^{\ast} =
x_2 + \varepsilon_2, \ x_3^{\ast} = x_3, \ u^{\ast} = u, \ v^{\ast}
= v, \ S_1^{\ast} = S_1, \ S_2^{\ast} = S_2,
\]
\[
\widetilde{T}_3: t^{\ast} = t, \ x_1^{\ast} = x_1, \ x_2^{\ast} =
x_2, \ x_3^{\ast} = x_3 + \varepsilon_3, \ u^{\ast} = u, \ v^{\ast}
= v, \ S_1^{\ast} = S_1, \ S_2^{\ast} = S_2,
\]
\[
\widetilde{T}_4: t^{\ast} = e^{2 \varepsilon_4} t, \ x_1^{\ast} =
e^{\varepsilon_4} x_1, \ x_2^{\ast} = e^{\varepsilon_4} x_2, \
x_3^{\ast} = e^{\varepsilon_4} x_3, \ u^{\ast} = u, \ v^{\ast} = v,
\ S_1^{\ast} = S_1, \ S_2^{\ast} = S_2,
\]
\[
\widetilde{T}_5: t^{\ast} = t, \ x_1^{\ast} = \theta_1(\varepsilon_5), \ x_2^{\ast} = \theta_2(\varepsilon_5), \ x_3^{\ast} = x_3, \ u^{\ast} = u, \ v^{\ast} = v, \ S_1^{\ast} = S_1, \ S_2^{\ast} = S_2,
\]
\[
\widetilde{T}_6: t^{\ast} = t + \varepsilon_6, \ x_1^{\ast} = \theta_1(\lambda \varepsilon_6), \ x_2^{\ast} = \theta_2(\lambda \varepsilon_6), \ x_3^{\ast} = x_3, \ u^{\ast} = u, \ v^{\ast} = v, \ S_1^{\ast} = S_1, \ S_2^{\ast} = S_2,
\]
\[
\widetilde{T}_7: t^{\ast} = e^{2 \varepsilon_7}t, \ x_1^{\ast} = e^{\varepsilon_7}\theta_1(\lambda \varepsilon_7), \ x_2^{\ast} = e^{\varepsilon_7}\theta_2(\lambda \varepsilon_7), \ x_3^{\ast} =e^{\varepsilon_7} x_3, \ u^{\ast} = u, \ v^{\ast} = v, \ S_1^{\ast} = S_1, \ S_2^{\ast} = S_2,
\]
where
\[
\theta_1(\tau) = x_1 \cos{\tau} + x_2 \sin{\tau}, \ \ \theta_2(\tau) = - x_1 \sin{\tau} + x_2 \cos{\tau}.
\]

{\bf Proof.} Let us  consider the case of the arbitrary function
$d_1(u)$ and $d_2(v)$.
According to Theorem 1 (see case 1 of
Table 1),  in this case the NHE system (\ref{14})--(\ref{15}) admits
 the eight-parameter group $G_8^{1}$ of invariance generated by
the groups $T_0, T_1, T_2, T_3, T_4$
 and  the rotation groups
\[
T_{{ab}}: \ t^{\ast} = t, \ x_a^{\ast} = - x_b
\sin\varepsilon_{ab} + x_a \cos\varepsilon_{ab}, \ x_b^{\ast} = x_b
\cos\varepsilon_{ab} + x_a \sin\varepsilon_{ab}, \ x_c^{\ast} = x_c,
\ u^{\ast} = u, \ v^{\ast} = v.
\]
where  $b < a = 1, 2, 3, \ c \neq
a, \ c \neq b$, and $\varepsilon_{ab}$ are the group parameters.

Since the BVP of the form (\ref{14})--(\ref{18}) has two free boundaries, $S_1(t,
\textbf{x}) = 0$ and $S_2(t, \textbf{x}) = 0$,  we need to extend
the group $G_8^{1}$ by adding identical transformations
for the new
variables $S_1^{\ast} = S_1$ and $S_2^{\ast} = S_2$. We will denote the obtained group by $\widetilde{G}_8^{1}$  (the relevant
sub-groups will be denoted in the same way,
 for instance, $\widetilde{T}_{{31}}$).

By straightforward calculations, one easily checks that   the boundary conditions (\ref{17})
and (\ref{18}) are invariant under  the group $\widetilde{G}_8^{1}$.
The situation is essentially different if one examines  the invariance of the boundary condition (\ref{16}) with
respect to $\widetilde{G}_8^{1}$.
To simplify  calculations,  the boundary conditions (\ref{16}) should be rewritten in the form
 (see the monograph
\cite{crank84}, P. 18)
\begin{equation}\label{pp1}
S_1(t, \textbf{x}) = 0: \ d_{1v} \nabla u \cdot \nabla S_1 = - H_v
\frac{\partial S_1}{\partial t} - \textbf{Q}(t) \cdot \nabla S_1, \
u = u_v
\end{equation}
 (we remind the reader that $|\nabla S_k| \neq 0$, $k = 1, 2$). It will be shown below  that the
 form of the  vector function $ \textbf{Q}(t) $ in  (\ref{pp1}) plays  a crucial role.

First of all, we  examine the invariance
with respect to the one-parameter Lie groups
forming the group $\widetilde{G}_8^{1}$.
Obviously, conditions (\ref{pp1})
are invariant with respect to the groups
$\widetilde{T}_1, \widetilde{T}_2$ and $\widetilde{T}_3$   for the  {\it arbitrary} smooth
 vector function $\textbf{Q}(t)$ .

 To be invariant under  the group $\widetilde{T}_0$,  the conditions must take place
\begin{equation}\label{pp1a}
\left. d_{1v} \nabla u^{\ast} \cdot \nabla S_1^{\ast} + H_v
\frac{\partial S_1^{\ast}}{\partial t^{\ast}} + \textbf{Q}(t^{\ast})
\cdot \nabla S_1^{\ast} \right \vert_{(39)}= 0, \ \left. u^{\ast} -
u_v \right \vert_{(39)} = 0,
\end{equation}
which lead to the requirement
\begin{equation}\label{pp2}
Q_a(t + \varepsilon_0) = Q_a(t), \ a = 1, 2, 3.
\end{equation}
Since (\ref{pp2}) must hold for arbitrary real values of   $t$ and $\varepsilon_T$, we conclude that
\begin{equation}\label{pp3}
Q_a(t) = q_a, \ a = 1, 2, 3,
\end{equation}
where $q_a$ are arbitrary constants.
Thus, the BVP of the form (\ref{14})--(\ref{18}) is invariant with respect to the
Lie group $\widetilde{T}_0$ if and only if conditions
(\ref{pp3}) take place.

In  a quite similar way,  one  examines the group
$\widetilde{T}_4$  and, as a result, arrives at the   requirement
\[
Q_a(t e^{2 \varepsilon_4}) e^{\varepsilon_4} = Q_a(t), \ a = 1, 2, 3
\]
what implies
\begin{equation}\label{pp4}
Q_a(t) = \frac{q_a}{\sqrt{t}}, \ a = 1, 2, 3.
\end{equation}
Thus, the BVP of the form (\ref{14})--(\ref{18}) is invariant with respect to the
Lie group $\widetilde{T}_4$ if and only if conditions
(\ref{pp4}) hold.

Prior to examine the invariance  of the boundary condition (\ref{16})
with respect to the group $\widetilde{T}_{{ab}}$,  we find how the
first derivations of the variables $u$ and $S_1$ are transformed:
\begin{equation}\label{pp5}
\frac{\partial u^{\ast}}{\partial x_a^{\ast}} = - \frac{\partial
u}{\partial x_b} \sin\varepsilon_{ab} + \frac{\partial u}{\partial
x_a} \cos\varepsilon_{ab}, \ \frac{\partial u^{\ast}}{\partial
x_b^{\ast}} = \frac{\partial u}{\partial x_b} \cos\varepsilon_{ab} +
\frac{\partial u}{\partial x_a} \sin\varepsilon_{ab}, \
\frac{\partial u^{\ast}}{\partial x_c^{\ast}} = \frac{\partial
u}{\partial x_c},
\end{equation}
%and
\begin{equation}\label{pp6}
\frac{\partial S_1^{\ast}}{\partial x_a^{\ast}} = - \frac{\partial
S_1}{\partial x_b} \sin\varepsilon_{ab} + \frac{\partial
S_1}{\partial x_a} \cos\varepsilon_{ab}, \ \frac{\partial
S_1^{\ast}}{\partial x_b^{\ast}} = \frac{\partial S_1}{\partial x_b}
\cos\varepsilon_{ab} + \frac{\partial S_1}{\partial x_a}
\sin\varepsilon_{ab}, \ \frac{\partial S_1^{\ast}}{\partial
x_c^{\ast}} = \frac{\partial S_1}{\partial x_c}, \ \frac{\partial
S_1^{\ast}}{\partial t^{\ast}} = \frac{\partial S_1}{\partial t}.
\end{equation}
Substituting (\ref{pp5}) and (\ref{pp6}) into (\ref{pp1a}) and making
relevant calculations, we arrive at the equality
\[
\left(Q_a(t) \cos\varepsilon_{ab} + Q_b(t) \sin\varepsilon_{ab}
\right) \frac{\partial S_1}{\partial x_a} + \left(- Q_a(t)
\sin\varepsilon_{ab} + Q_b(t) \cos\varepsilon_{ab} \right)
\frac{\partial S_1}{\partial x_b} + Q_c(t) \frac{\partial
S_1}{\partial x_c} = \textbf{Q}(t) \cdot \nabla S_1,
\]
leading to the algebraic equations
\[
Q_a(t) \cos\varepsilon_{ab} + Q_b(t) \sin\varepsilon_{ab} = Q_a(t),\
Q_a(t) \sin\varepsilon_{ab} - Q_b(t) \cos\varepsilon_{ab} = -Q_b(t).
\]
Since  $\varepsilon_{ab}$ is an arbitrary parameter, we immediately    conclude that
\begin{equation}\label{pp7}
Q_a(t) = Q_b(t) \equiv 0.
\end{equation}
It means that the BVP of the form (\ref{14})--(\ref{18}) is invariant with respect to the
Lie group $\widetilde{T}_{{ab}}$ if and only if conditions
(\ref{pp7}) hold, while the function $Q_c(t)$ is an arbitrary smooth
function, i.e.  the  vector function $ \textbf{Q}(t) $ contains only one non-vanish  component.

 Thus, we proved that the BVP of the form (\ref{14})--(\ref{18}) is invariant under  $\widetilde{T}_0,
\widetilde{T}_4$ and
$\widetilde{T}_{{ab}} $ iff  the  vector function $ \textbf{Q}(t)
$ satisfies the restrictions (\ref{pp3}), (\ref{pp4}) and
(\ref{pp7}), respectively.
 To finish the examination of the BVP with  arbitrary functions
$d_1(u)$ and $d_2(v)$,
 we need to investigate the case
 of the arbitrary   one-parameter Lie
group from the group $\tilde{g}_5 = (\widetilde{T}_0,
\widetilde{T}_4, \widetilde{T}_{{21}}, \widetilde{T}_{{31}},
\widetilde{T}_{{32}})$. According to the general Lie theory,
each  one-parameter group of the group $\tilde{g}_5$ corresponds to
a  linear combination of the generators $P_t$, $D_0$ and $J_{ab}$ of the form
%%\ b < a = 1, 2, 3$
\[
L = \lambda_1 J_{21} + \lambda_2 J_{31} + \lambda_3 J_{32} +
\lambda_4 P_t + \lambda_5 D_0, \ \lambda_1^2 + \ldots + \lambda_5^2
\neq 0,
\]
where $\lambda_1, \ldots, \lambda_5$ are the given real constants.

It should be stressed that the form of the operator $L$ can be simplified by using the transformations of variables $x_a, \ a = 1, 2, 3$ from the group $\widetilde{E}_{\mathrm{eq}}^{\mathrm{BVP}}$, namely the linear combination $\lambda_1 J_{21} + \lambda_2 J_{31} + \lambda_3 J_{32}$ may be reduced to a single operator of rotation, for example, to the operator $J_{12}$. Indeed,  using the transformations
\[
t \rightarrow t, \ X \rightarrow A_1(\beta_1) A_2(\beta_2) A_3(\beta_3) X
\]
where   the rotation angles $\beta_a, \ a =1, 2, 3$  satisfy  the conditions
\[
\lambda_1 \cos\beta_2 \sin\beta_3 - \lambda_2 \left(\cos\beta_1 \cos\beta_3 - \sin\beta_1 \sin\beta_2 \sin\beta_3 \right) - \lambda_3 \left( \sin\beta_1 \cos\beta_3 + \cos\beta_1 \sin\beta_2 \sin\beta_3 \right) = 0,
\]
\[
\lambda_1 \sin\beta_2 - \lambda_2 \sin\beta_1 \cos\beta_2 + \lambda_3 \cos\beta_1 \cos\beta_2 = 0.
\]
one simplifies  the operator $L$ to the form
\[
L = \lambda J_{21} + \lambda_4 P_t + \lambda_5 D_0, \ \lambda^2 + \lambda_4^2 + \lambda_5^2 \neq 0,
\]
where
\[
\lambda = \lambda_1 \cos\beta_2 \cos\beta_3 + \lambda_2 \left(\cos\beta_1 \sin\beta_3 + \sin\beta_1 \sin\beta_2 \cos\beta_3 \right) + \lambda_3 \left( \sin\beta_1 \sin\beta_3 - \cos\beta_1 \sin\beta_2 \cos\beta_3 \right).
\]

Now one can easily see that there are only two cases
(depending on the parameters $\lambda, \lambda_4$ and $ \lambda_5$ !),
leading to new  one-parameter   invariance groups of the  BVP of the form (\ref{14})--(\ref{18})
with the correctly specified  vector function $ \textbf{Q}(t) $.
This occurs iff: {\it a)} $\lambda \lambda_4 \neq 0$ and $\lambda_5 = 0$, and {\it b)} $\lambda \lambda_5 \neq 0$.

Let us consider the most troublesome case {\it b)}.  Without loss of generality, one can
assume $\lambda_4 = 0$ and $\lambda_5 = 1$, therefore, the corresponding
one-parameter Lie  group  is  $\widetilde{T}_7$ (see Remark~3). Substituting transformations from this group into invariance conditions (\ref{pp1a}), we obtain  the system
\[
\widehat{Q}_1 \cos(\lambda\varepsilon_7) - \widehat{Q}_2 \sin(\lambda\varepsilon_7) = Q_1(t),
\]
\[
\widehat{Q}_1 \sin(\lambda\varepsilon_7) + \widehat{Q}_2 \cos(\lambda\varepsilon_7) = Q_2(t),
\]
\[
\widehat{Q}_3 = Q_3(t),
\]
which can be  rewritten as follows:
\[
\widehat{Q}_1 = Q_1(t) \cos(\lambda\varepsilon_7) + Q_2(t) \sin(\lambda\varepsilon_7),
\]
\[
\widehat{Q}_2 = - Q_1(t) \sin(\lambda\varepsilon_7) + Q_2(t) \cos(\lambda\varepsilon_7),
\]
\[\widehat{Q}_3 = Q_3(t),
\]
where $\widehat{Q}_a = Q_a(e^{2 \varepsilon_7}t)e^{\varepsilon_7}, \ a = 1, 2, 3$. So  unknown components of the
vector function $\textbf{Q}(t)$  can be found from the  system of
functional equations obtained  above:
\begin{equation}\label{f1}
Q_1(t) = \frac{q_1 \cos\tau + q_2 \sin\tau}{\sqrt t}, \ Q_2(t) = \frac{- q_1 \sin\tau + q_2 \cos\tau}{\sqrt t}, \ Q_3(t) = \frac{q_3}{\sqrt t},
\end{equation}
where $\tau = \frac{1}{2} \lambda \log t$, $q_a, \ a = 1, 2, 3$ are arbitrary real constants obeying the condition $q_1 q_2\lambda \neq 0$.

Hence, the BVP of the form (\ref{14})--(\ref{18}) is invariant with respect to the Lie group $\widetilde{T}_{7}$ if and only if the conditions (\ref{f1}) hold. Case {\it b)} has been completely investigated.

Case {\it a)} has been examined  in a quite similar way. As result, we have proved  that the BVP of the form (\ref{14})--(\ref{18}) is invariant with respect to the Lie group $\widetilde{T}_{6}$ iff the conditions
\[
Q_1(t) = q_1 \cos(\lambda t) + q_2 \sin(\lambda t), \ Q_2(t) = - q_1 \sin(\lambda t) + q_2 \cos(\lambda t), \ Q_3(t) = q_3, \ \ q_1 q_2 \lambda \neq 0
\]
take place.

Thus, we  conclude
from the analysis  carried out above that
the trivial group of invariance
$\widetilde{G}^{0}$ of the BVP of the form (\ref{14})--(\ref{18}) is the
three-parameter Lie group, generated by the groups $\widetilde{T}_1,
\widetilde{T}_2$ and $\widetilde{T}_3$ and one is listed in case 1 of Table \ref{tab2}.  $\widetilde{G}^{0}$  is extended either to
four-parameter  group (cases  2--4  of  Table \ref{tab2}) or  five-parameter  group (cases  5 and 6  of  Table \ref{tab2})  if and only if the relevant conditions at the
vector function $\textbf{Q}(t)$ take place.  These conditions  have been derived
by comparing  the components of $\textbf{Q}(t)$ from formulae
 (\ref{pp3}), (\ref{pp4}) and (\ref{pp7}) and those from the examination of cases {\it a)} and {\it b)}.
One  easily notes  that there are no other possibilities to extend  the group  $\widetilde{G}^{0}$.
 It  means that  the case of the  arbitrary functions $d_1(u)$ and $d_2(v)$  is
completely  investigated.

The next step of the proof is to show that cases 2--9 from Table
\ref{tab1} do not lead to any new symmetries of the BVP of the form
(\ref{14})--(\ref{18}). Consider case 2 when $d_1(u) = k_1$ and
$d_2(v)$ is an arbitrary function. According to Theorem 1, the NHE system
(\ref{14})--(\ref{15}) admits the infinite-dimensional  Lie algebra  formed by the operators from the basic algebra $AE(1,3)$
and the operators $u \partial_u,$ $\alpha(t,\textbf{x})\partial_u$.
The corresponding groups to these operators are generated by the
transformations
\[
\widetilde{T}_U: \ t^{\ast} = t, \ x^{\ast}_a = x_a, a = 1, 2, 3, \ u^{\ast} =
e^{\varepsilon_U} u, \ v^{\ast} = v, \ S_1^{\ast} = S_1, \
S_2^{\ast} = S_2,
\]
\[
\widetilde{T}_{\infty}: \ t^{\ast} = t, \ x^{\ast}_a = x_a, a = 1, 2, 3, \
u^{\ast} = u + \alpha(t,\textbf{x}) \varepsilon_{\infty}, \ v^{\ast}
= v, \ S_1^{\ast} = S_1, \ S_2^{\ast} = S_2,
\]
Let us prove that any BVP in question is not invariant with respect to
the group $\widetilde{T}_U$.
In fact, the second  invariance
condition  in   (\ref{pp1a}) takes the form
\[
e^{\varepsilon_U} u_v = u_v
\]
and can be satisfied  for an arbitrary value of the parameter
$\varepsilon_U$ only in the case  $u_v = 0$.
On the other hand,  if  $u_v = 0$ then $u_m \neq 0$ and, thereby, the second
boundary condition in  (\ref{17}) is not invariant with respect to
$\widetilde{T}_U$. In a similar manner, it can be shown that  any BVP in question is not invariant with respect to
$\widetilde{T}_{\infty}$.

Now,
we must examine each
one-parameter group, corresponding to a  linear combination of the
operators
$P_t$, $D_0$, $J_{ab}, \ b < a = 1,2,3$, $u \partial_u$
and $\alpha(t,\textbf{x})\partial_u$ (see  case 2 of Table
\ref{tab1}):
\[
Y = \lambda_1 J_{21} + \lambda_2 J_{31} + \lambda_3 J_{32} +
\lambda_4 P_t + \lambda_5 D_0 + \lambda_6 u\partial_u + \lambda_7
 \alpha(t,\textbf{x})\partial_u, \ \lambda_6^2 + \lambda_7^2 \neq 0,
\]
where $\lambda_1, \ldots, \lambda_7$ are arbitrary real constants.
To avoid cumbersome formulae, we  present  in an explicit form only
the point transformations for  the variable $u$:
\begin{equation}\label{pp8}
u^{\ast} = e^{\lambda_6 \varepsilon_{Y}} u + \lambda_7
\int_0^{\varepsilon_{Y}} \alpha^{\ast}(\tau) e^{\lambda_6 (\tau -
\varepsilon_{Y})} d \tau,
\end{equation}
where $\alpha^{\ast}(\tau) = \alpha \left(t^{\ast}(\tau),
\textbf{x}^{\ast}(\tau)\right)$. Obviously, if $\lambda_6 = 0, \lambda_7 \neq 0$ or $\lambda_6 \neq 0, \lambda_7 = 0$, then
the BVP under study is not invariant with respect to the relevant
one-parameter Lie group because one  uses the result obtained above.
 Hence, we  should  examine only the case
$\lambda_6 \lambda_7 \neq 0$. Taking into account formula
(\ref{pp8}), the second invariance condition  from  (\ref{pp1a}) takes the form
\[
\lambda_7 \int_0^{\varepsilon_{Y}} \alpha^{\ast}(\tau) e^{\lambda_6
(\tau - \varepsilon_{Y})} d \tau = u_v \left(1 - e^{\lambda_6
\varepsilon_{Y}}\right).
\]
Thus, we obtain
\[
u^{\ast} = e^{\lambda_6 \varepsilon_{Y}} u + u_v \left(1 -
e^{\lambda_6 \varepsilon_{Y}}\right).
\]
Using  this formula, the invariance criterion for
the second  boundary condition from  (\ref{17}) leads to the condition
\[
u_v \left(1 - e^{\lambda_6 \varepsilon_{Y}}\right) = u_m \left(1 -
e^{\lambda_6 \varepsilon_{Y}}\right).
\]
Since this equality must hold for arbitrary values of the group
parameter $\varepsilon_{Y}$,
we immediately obtain that $u_m = u_v$.
It is nothing other than the contradiction because $u_m \neq u_v$
and $\lambda_6 \neq 0$. Thus, we  conclude that the operator $Y$ is
not  a Lie symmetry operator for any BVP from class
(\ref{14})--(\ref{18}). Case 2 from Table \ref{tab2} is
completely examined. Obviously, case 3 from Table \ref{tab1}  can
be studied in a similar manner to case 2 (the boundary
conditions (\ref{17}) and (\ref{18}) should be used) and the same
result is obtained.

Consider cases  4---8 from Table \ref{tab1} when   $d_1(u)$ and
$d_2(v)$  are  specified  non-constant  functions.
It  turns out  that cases
4---7 from Table \ref{tab1} can be  examined in a similar manner as we
did it  in case 2 using  the group $\widetilde{T}_U$. Finally, no new Lie groups of invariance are obtained.

The most non-trivial is  the case of conformal power $n = m = -
\frac{4}{5}$ (see case 8 in Table \ref{tab1}) because of the
conformal operators $K_b = |\textbf{x}|^2 \partial_{x_b} - 2 x_b x_a
\partial_{x_a} + 5 x_b u \partial _u + 5 x_b v \partial_v$, $b < a = 1, 2, 3,$  which generate
the  one-parameter  Lie groups
\begin{eqnarray}
& & T_{K_b}: \ x_a^{\ast} = \frac{x_a}{1 - 2 x_b \varepsilon_{K_b} +
|\textbf{x}|^2 \varepsilon_{K_b}^2}, \ x_b^{\ast} = \frac{x_b -
|\textbf{x}|^2 \varepsilon_{K_b}}{1 - 2 x_b \varepsilon_{K_b} +
|\textbf{x}|^2 \varepsilon_{K_b}^2}, \ x_c^{\ast} = \frac{x_c}{1 - 2
x_b \varepsilon_{K_b} + |\textbf{x}|^2 \varepsilon_{K_b}^2}, \nonumber \\
& & \qquad \quad t^{\ast} = t, \ u^{\ast} = u \left(1 - 2 x_b
\varepsilon_{K_b} + |\textbf{x}|^2 \varepsilon_{K_b}^2
\right)^{\frac{5}{2}}, \ v^{\ast} = v \left(1 - 2 x_b
\varepsilon_{K_b} + |\textbf{x}|^2 \varepsilon_{K_b}^2
\right)^{\frac{5}{2}}, \nonumber
\end{eqnarray}
where $a, b, c = 1, 2, 3; \ a \neq b, a \neq c, b \neq c$.
One notes that  the boundary
condition (\ref{18})  is not invariant under   $T_{K_b}$. In fact,
 the  invariance conditions
\begin{equation}\label{pp9}
\lim_{|\textbf{x}| \rightarrow + \infty} |\textbf{x}^{\ast}| = +
\infty, \ \left. v - v_{\infty} \right \vert_{(\ref{18})} = 0.
\end{equation}
are not satisfied because
\[
\lim_{|\textbf{x}| \rightarrow + \infty} |\textbf{x}^{\ast}| = \lim_{|\textbf{x}| \rightarrow + \infty}
\frac{|\textbf{x}|}{\left(1 - 2 x_b
\varepsilon_{K_b} + |\textbf{x}|^2
\varepsilon_{K_b}^2\right)^{\frac{1}{2}}}=
\frac{1}{\varepsilon_{K_b}} \not = +
\infty.
\]
Thus,we conclude that  any BVP of the form (\ref{14})--(\ref{18}) is
not invariant with respect to the Lie  group $\widetilde{T}_{K_b}$.

Finally, we have examined all possible
 one-parameter
groups, corresponding to  linear combinations  of the operators
$K_b, \ b = 1, 2, 3$ and $D$ (see case 8 in Table \ref{tab1})
\[
Z = \lambda_1 K_1 + \lambda_2 K_2 + \lambda_3 K_3 + \lambda_4 D, \
\lambda_1^2 + \lambda_2^2 + \lambda_3^2 \neq 0,
\]
 and shown that  the boundary
condition (\ref{18})  is not invariant under  those groups.

Nevertheless, the case of linear governing system
(\ref{14})--(\ref{15}) (case 9 in Table \ref{tab2}) has a cumbersome
algebra of invariance; it was  examined in a quite similar way
to case 2 and  no new  invariance groups found.

The proof is now complete. $\blacksquare$

\begin{remark}
In Theorem 2, the special  case of system (\ref{14})--(\ref{15}) with $d_1(u)=d_2(v)$ (see case 10 in Table \ref{tab1}) was not examined  because the relevant BVP with the equal diffusivities is  unrealistic from the physical point of view. In fact, the vast majority of substances have different physical characteristics of solid, liquid, and gas phases.
\end{remark}

\section{\bf Symmetry reduction and invariant solutions of BVPs from class (\ref{14})--(\ref{18}) with the constant energy flux}

\subsection{\bf Optimal system of subalgebras of the invariance  algebra}

Let us consider a nonlinear  model of heat transfer processes
   in metals under the action of intense constant energy flaxes directed perpendicular to the metal  surface.
     This model coincides with the BVP (\ref{8})--(\ref{12}) (see also   (\ref{14})--(\ref{18})), where $\textbf{Q}(t) = \textbf{q} \equiv (0, 0, q), \ q = \mbox{const}$. According to Theorem 2 such a BVP  admits the  five-parameter Lie group $\widetilde{G}_5$ of invariance (see case 5, in Table \ref{tab2}) formed by the one-parameter groups $\widetilde{T}_0, \widetilde{T}_1, \widetilde{T}_2, \widetilde{T}_3$ and $\widetilde{T}_{{12}}$. This group corresponds to the five-dimensional Lie algebra $A_5$ with the basic operators
\[
P_t = \frac{\partial}{\partial t}, \ P_a = \frac{\partial}{\partial
x_a}, \ J_{12} = x_2 P_1 - x_1 P_2, \, a = 1, 2, 3.
\]

Our primary aim  is to show how using these operators one  can reduce the BVP  (\ref{14})--(\ref{18}), where $\textbf{Q}(t) = \textbf{q}$,  to BVPs  of lower  dimensions.
  We use for this purpose the  \emph{optimal systems of $s$-dimensional subalgebras} ($s\leq5$) of $A_5$.
  All such subalgebras are non-conjugate  up to  the group of inner automorphisms of the group $\widetilde{G}_5$.
  To construct a full list   of the optimal systems, we  represent   the algebra $A_5$ as follows:
$A_5 = \left<P_1, P_2, J_{12}\right> \oplus \left<P_3\right> \oplus \left<P_t\right>$.
  Now by using the well-known Lie-Goursat classification method for the subalgebras of algebraic sums of Lie algebras \cite{pathera-1} (see the monograph \cite{bar} for details) and the results of the subalgebras classification of low-dimensional real Lie algebras \cite{pathera-2},  it is easy to obtain the complete list of subalgebras of the algebra $A_5$. This list can be divided into subalgebras of different dimensionality.

 One-dimensional subalgebras:
\[
 \left<P_3 \cos\phi + P_t \sin\phi \right>, \ \left<P_1 + \alpha\left(P_3 \cos\phi + P_t \sin\phi \right)\right>, \ \left<J_{12} + \beta \left(P_3 \cos\phi + P_t \sin\phi \right)\right>;
\]

 Two-dimensional subalgebras:
\[
\left<P_3, P_t\right>, \ \left<P_1 + \alpha \left(P_3 \cos\phi + P_t \sin\phi \right), P_2\right>, \ \left<P_1 + \alpha \left(P_3 \cos\phi + P_t \sin\phi \right), P_3 \sin\phi - P_t \cos\phi\right>,
\]
\[
\left<J_{12} + \beta \left(P_3 \cos\phi + P_t \sin\phi \right), P_3 \sin\phi - P_t \cos\phi\right>;
\]

Three-dimensional subalgebras:
\[
\left<P_1, P_3, P_t\right>, \ \left<J_{12}, P_3, P_t\right>, \ \left<P_1 + \alpha \left(P_3 \cos\phi + P_t \sin\phi \right), P_2, P_3 \sin\phi - P_t \cos\phi\right>,
\]
\[
\left<J_{12} + \beta \left(P_3 \cos\phi + P_t \sin\phi \right), P_1, P_2\right>;
\]

Four-dimensional subalgebras :
\[
\left<P_1, P_2, P_3, P_t\right>, \ \left<J_{12} + \beta \left(P_3 \cos\phi + P_t \sin\phi \right), P_1, P_2, P_3 \sin\phi - P_t \cos\phi\right>;
\]

Five-dimensional subalgebra:
\[
\left<J_{12}, P_1, P_2, P_3, P_t\right>.
\]
where $ \alpha \geq 0$ and $\beta$ are arbitrary real constants, $0 \leq \phi < \pi$.

We  remaind the reader  that  the additional conditions on the functions $S_k(t, \textbf{x})$, $\textbf{V}_k(t, \textbf{x})$ and $\textbf{Q}(t)$  arising in  the BVP class (\ref{14})--(\ref{18}) have been imposed, namely
\begin{equation}\label{4.1}
\frac{\partial S_k}{\partial t} \neq 0, \ |\nabla S_k| \neq 0, \ \textbf{Q}(t) \cdot \textbf{n}_1 \neq 0, \ \textbf{V}_k \cdot \textbf{n}_k \neq 0, \  k=1,2.
\end{equation}
On the other hand, the Lie-Goursat classification method is a purely algebraic procedure; therefore,  some subalgebras
  presented above  lead to invariant solutions,
    which do not satisfy the restrictions (\ref{4.1}) . For example, the algebra $\left<J_{12}, P_3, P_t\right>$  generates  the ansatz
\[
u = u(r), \ v = v(r), \ S_k = S_k(r), \ k = 1, 2,
\]
where $r = \sqrt{x_1^2 + x_2^2}$. Obviously, one sees that  $\frac{\partial S_k}{\partial t} = 0$ and $\textbf{q} \cdot \textbf{n}_1 = 0$. Thus, the contradiction is obtained and we conclude that the algebra $\left<J_{12}, P_3, P_t\right>$ leads to  solutions, which do not have any physical meaning.

The complete list of the subalgebras, leading to invariant solutions of the BVPs in question satisfying the restrictions (\ref{4.1}), reads as follows.

 One-dimensional subalgebras:
\[
 \left<P_3 \cos\phi + P_t \sin\phi \right> \left(\phi \neq 0, \frac{\pi}{2}\right), \ \left<P_1 + \alpha\left(P_3 \cos\phi + P_t \sin\phi \right)\right>, \ \left<J_{12} + \beta \left(P_3 \cos\phi + P_t \sin\phi \right)\right>;
\]

 Two-dimensional subalgebras:
\[
\left<P_1 + \alpha \left(P_3 \cos\phi + P_t \sin\phi \right), P_2\right>, \ \left<P_1 + \alpha \left(P_3 \cos\phi + P_t \sin\phi \right), P_3 \sin\phi - P_t \cos\phi\right> \left(\phi \neq 0, \frac{\pi}{2}\right),
\]
\[
\left<J_{12} + \beta \left(P_3 \cos\phi + P_t \sin\phi \right), P_3 \sin\phi - P_t \cos\phi\right> \left(\phi \neq 0, \frac{\pi}{2}\right);
\]

Three-dimensional subalgebras:
\begin{equation}\label{4.0a}
\left<P_1 + \alpha \left(P_3 \cos\phi + P_t \sin\phi \right), P_2, P_3 \sin\phi - P_t \cos\phi\right> \left(\phi \neq 0, \frac{\pi}{2}\right),
\end{equation}
\begin{equation}\label{4.0b}
\left<J_{12} + \beta \left(P_3 \cos\phi + P_t \sin\phi \right), P_1, P_2\right> \left(\phi \neq 0, \frac{\pi}{2} \ \mbox{if} \ \beta \neq 0\right);
\end{equation}

Four-dimensional subalgebras :
\begin{equation}\label{4.0c}
\left<J_{12}, P_1, P_2, P_3 \sin\phi - P_t \cos\phi\right> \left(\phi \neq 0, \frac{\pi}{2}\right),
\end{equation}
where $ \alpha \geq 0$ and $\beta$ are arbitrary real constants, $0 \leq \phi < \pi$.

\subsection{Symmetry reduction and invariant solutions }

Now, one may use each algebra from this list for reducing the BVP of the form (\ref{14})--(\ref{18})
with  $\textbf{Q}(t) = \textbf{q}$
  to the BVP of
lower dimensionality and to construct the exact solutions of the problem obtained.
We also note that   three- and four-dimensional subalgebras  (\ref{4.0b}) and (\ref{4.0c})  generate the same ansatz
to find the unknown functions $u, \, v, \,  S_1$  and $S_2$:
\begin{equation}\label{4.2*}
u = u(z), \ v = v(z), \ S_k = S_k(z), \ k = 1, 2, \, z = x_3 - \mu t.
\end{equation}
 (Hereafter, we use the designation $\mu = - \tan \phi$).
 The application of  three-dimensional subalgebra  (\ref{4.0a}) also yields ansatz (\ref{4.2*}) but with the invariant  variable $z = \alpha^* x_1+ x_3 - \mu t,   \, \alpha^* \in \mathbb{R}$.
 Ansatz   (\ref{4.2*}) reduces the BVP of the form (\ref{14})--(\ref{18}) to the BVP  for ODEs, which was studied in details in our earlier papers \cite {ch-od90,ch93,ch-kov-11}. This ansatz leads to the  plane wave solutions
 of the BVP  in question, while melting and evaporation surfaces are two parallel
 planes moving with  unknown  velocity $\mu$ along the axes $0 x_3$.

New non-trivial reductions occur if one applies  one- and two-dimensional subalgebras. Let us consider the algebra $\left<J_{12} + \beta \left(P_3 \cos\phi + P_t \sin\phi \right), P_3 \sin\phi - P_t \cos\phi\right>$. Solving the corresponding system of characteristic equations, one obtains the ansatz
\begin{equation}\label{4.2}
u = u(r,z), \ v = v(r,z), \ S_k = S_k(r,z), \ k = 1, 2,
\end{equation}
where $z = x_3 - \mu t - \beta \arctan \frac{x_1}{x_2}, \ r = \sqrt{x_1^2 + x_2^2}$ are the invariant variables.

Substituting ansatz (\ref{4.2}) into the BVP of the form (\ref{14})--(\ref{18}) with  $\textbf{Q}(t) = \textbf{q}$  and making the relevant calculations, we arrive at the two-dimensional BVP
\begin{eqnarray}
& & \frac{1}{r} \frac{\partial}{\partial r} \left(r d_1(u) \frac{\partial u}{\partial r} \right) + \left(\frac{\beta^2}{r^2} + 1 \right)\frac{\partial}{\partial z} \left(d_1(u) \frac{\partial u}{\partial z} \right) + \mu \frac{\partial u}{\partial z} = 0,\label{4.3} \\
& & \frac{1}{r} \frac{\partial}{\partial r} \left(r d_2(v) \frac{\partial v}{\partial r} \right) + \left(\frac{\beta^2}{r^2} + 1 \right)\frac{\partial}{\partial z} \left(d_2(v) \frac{\partial v}{\partial z} \right) + \mu \frac{\partial v}{\partial z} = 0,\label{4.4} \\
 & & \qquad S_{1}(r,z) = 0:\ d_{1v}
\nabla'u \cdot \nabla'S_1 = \left (\mu H_v - q \right) \frac{\partial S_1}{\partial z}, \ u = u_v,\label{4.5} \\
& & \qquad S_{2}(r,z) = 0: \ d_{2m} \nabla'v \cdot \nabla'S_2 = d_{1m} \nabla'u \cdot \nabla'S_2 + \mu H_m \frac{\partial S_2}{\partial z},\ u = u_m, \ v =
v_m,\label{4.6}
\\ & & \qquad r^2 + z^2 = +\infty: \ v = v_{\infty},\label{4.7}
\end{eqnarray}
where $\mu$ is an unknown parameter and
  $\nabla'= \left(\frac{\partial}{\partial r}, \sqrt{\frac{\beta^2}{r^2} + 1}\frac{\partial}{\partial z} \right)$.

  Nevertheless, the BVP of the form (\ref{4.3})--(\ref{4.7}) is much simple object than the initial  BVP, but
   it is still the nonlinear problem with the basic two-dimensional PDEs.
   Our purpose is to reduce one to a BVP with basic ODEs.
    Of course, one may apply different technics to realize such reduction; however, we confine ourselves to the case when the invariant variables $r$ and $z$ admit clear physical meaning. It happens  for  $\beta = 0$ because then
  % an interesting example. Let
   the first variable $z$  makes the transition to a moving coordinate system (in the direction of the variable  $x_3$)   with the origin at the evaporation surface,  while the variable $r$  presents   the radial symmetry of the process with respect to the variables $x_1$ and $x_2$. Obviously, such a situation   takes  place if
the surface  bounded by a circle of the radius $R$ is exposed by the flux $\textbf{Q}(t) = \textbf{q}$.

  Thus, setting  $\beta = 0$,  we may  consider the  ansatz
\begin{equation}\label{4.8}
u = u(\omega), \ v = v(\omega), S_k = S_k(\omega), \ \omega = z + \sqrt{z^2 + r^2}, \ k = 1, 2,
\end{equation}
used earlier  in  \cite{ch-2003} for a similar purposes. Note that it is a non-Lie ansatz because  the  maximal  algebra  of invariance  of system
 (\ref{4.3})--(\ref{4.4}) (with arbitrary non-constant functions $d_1(u) $ and $d_2(v)$)
 is trivial  and  generated by the  operator $ \frac{\partial }{\partial z}$.

\begin{remark} Recently we found that ansatz (\ref{4.8}) with $\omega$ determined from the cubic equation
\begin{equation}\label{4.8**}
1- \frac{2}{\omega}(x_3-\mu t)= \frac{x_1^2}{\omega^2} + \frac{ x_2^2}{\omega(\omega+\kappa)}, \,  \kappa \in \mathbb{R}
\end{equation}
was used in  paper   \cite{lyubov} to construct the exact solution of the BVP of the form (\ref{14})--(\ref{18}) with  $\textbf{Q}(t) = \textbf{q}$ and the constant  diffusivities $d_1(u) $ and  $d_2(v)$. Setting  $ \kappa =0$ and $x_3-\mu t=z$  in  (\ref{4.8**}) one arrives at ansatz  (\ref{4.8}).  However, we have  checked  that ansatz  (\ref{4.8})  with  $\omega$  defined from  (\ref{4.8**}) ( with any non-zero  $ \kappa$ !) is not applicable to reduce the BVP of the form (\ref{14})--(\ref{18}) with non-constant diffusivity.
\end{remark}

Substituting  ansatz (\ref{4.8}) into the BVP of the form (\ref{4.3})--(\ref{4.7}) and taking into account the relations
\[
\nabla'u \cdot \nabla'S_k = \frac{2 \omega}{\sqrt{z^2 + r^2}} \frac{d u}{d \omega} \frac{d S_k}{d \omega}, \ \nabla'v \cdot \nabla'S_k = \frac{2 \omega}{\sqrt{z^2 + r^2}} \frac{d v}{d \omega} \frac{d S_k}{d \omega}, \ \frac{\partial S_k}{\partial z} = \frac{\omega}{\sqrt{z^2 + r^2}} \frac{d S_k}{d \omega}, \ k = 1, 2,
\]
we obtain the BVP for ODEs:
\begin{eqnarray}
& & \frac{d}{d \omega} \left(\omega d_1(u) \frac{d u}{d \omega} \right) + \mu \frac{\omega}{2} \frac{d u}{d \omega} = 0, \ \ 0 < \omega_1 < \omega < \omega_2,\label{4.9} \\
& & \frac{d}{d \omega} \left (\omega d_2(v) \frac{d v}{d \omega} \right) + \mu \frac{\omega}{2} \frac{d v}{d \omega} = 0, \ \ \omega > \omega_2,\label{4.10} \\ & & \qquad \omega = \omega_1: \ 2 d_{1v}
\frac{d u}{d \omega} = \mu H_v - q, \ u = u_v,\label{4.11} \\
& & \qquad \omega = \omega_2: \ 2 d_{2m} \frac{d v}{d \omega} = 2 d_{1m} \frac{d u}{d \omega} + \mu H_m,\ u = u_m, \ v = v_m,\label{4.12}
\\ & & \qquad \omega = +\infty: \ v = v_{\infty},\label{4.13}
\end{eqnarray}
where $\omega_k, \ k = 1, 2$ and $\mu$ are parameters to be determined.

Now, we can define the form of the free surfaces $S_k(t, \textbf{x}) = 0, \ k = 1, 2$ because in accordance with ansatz (\ref{4.8})
\[
S_k(\omega) \equiv z + \sqrt{z^2 + r^2} = \omega_k, \  k = 1, 2.
\]
Obviously,
the last equations can be rewritten in  the  form
\begin{equation}\label{4.8*}
\frac{x_1^2 + x_2^2}{\omega_k^2} = 1 - \frac{2 z}{\omega_k}, \ k = 1, 2.
\end{equation}
Thus, the equations obtained define some paraboloids of revolution in the space of variables $x_1, x_2, z$ (see Fig.1).  From the physical point of view, unknown parameters should satisfy the inequalities
  $\omega_2 > \omega_1 > 0$. Moreover, the parameter $ \omega_1 $  can be defined  as follows.  If one sets  $z=0$  in  (\ref{4.8*}) then $\omega_1= \sqrt {x_1^2 + x_2^2}$. On the other hand,
 a part of  the surface  bounded by a circle of the radius $R$ is only exposed by
the flux $\textbf{Q}(t) = \textbf{q}$, i.e. we can set   $\omega_1=R$ without losing the generality.

Let us  now turn to the construction of the exact solutions of problem (\ref{4.9})--(\ref{4.13}).
 In fact, the general solution of the  nonlinear equations  (\ref{4.9}) and (\ref{4.10}) is unknown.  However, in some  cases,  one  is  known (see, e.g., \cite{pol-za}). Here, we consider two  cases in details.
 % when system (\ref{4.9}) and (\ref{4.10}) can be exactly solved.

\medskip

{\textbf{Example 1.}} \emph{The BVP of the form (\ref{4.9})--(\ref{4.13}) with $d_1(u) = a_1$ and $d_2(v) = a_2$, where $a_1, a_2 \in \mathbb{R}^{+}$.}

In this case, the general solutions of equations (\ref{4.9}) and (\ref{4.10}) are given in an explicit form by the formulae
\begin{equation}\label{4.14}
u = C_1 \Phi_1(\omega) + C_2, \ v = C_3 \Phi_2(\omega) + C_4,
\end{equation}
where $\Phi_k(\omega) = \int_{\omega}^{+ \infty} \omega^{-1} e^{- \frac{\mu}{2 a_k} \omega} d\omega, \ k = 1, 2$, $C_1, \ldots, C_4$ are  arbitrary to-be-determined constants.
Substituting solution (\ref{4.14}) into the boundary conditions (\ref{4.11})--(\ref{4.13}) and taking into account the formulae
\[
\frac{d \Phi_k}{d \omega} = \omega^{-1} e^{- \frac{\mu}{2 a_k} \omega}, \ k = 1, 2,
\]
we arrive at the exact solution
\begin{equation}\label{4.15}
u = \frac{u_v - u_m}{\Phi_1(R) - \Phi_1(\omega_2)} \Phi_1(\omega) + \frac{u_m \Phi_1(R) - u_v \Phi_1(\omega_2)}{\Phi_1(R) - \Phi_1(\omega_2)}, \ v = \frac{v_m - v_{\infty}}{\Phi_2(\omega_2)} \Phi_2(\omega) + v_{\infty},
\end{equation}
where the parameters $\omega_2$ and $\mu$  must be found from  the transcendent equation system
\begin{eqnarray}
& & 2 d_{1 v} \frac{u_v - u_m}{\Phi_1(R) - \Phi_1(\omega_2)} R^{- 1} e^{- \frac{\mu}{2 a_1} R} = \mu H_v - q,\nonumber \\ & & 2 d_{2 m} \frac{v_m - v_{\infty}}{\Phi_2(\omega_2)} \omega_2^{-1} e^{- \frac{\mu}{2 a_2} \omega_2} = 2 d_{1 m} \frac{u_v - u_m}{\Phi_1(R) - \Phi_1(\omega_2)} \omega_2^{-1} e^{- \frac{\mu}{2 a_1} \omega_2} + \mu H_m,\nonumber
\end{eqnarray}

Finally, using formulae (\ref{4.15}) , (\ref{4.2}) and (\ref{4.8}), we obtain the exact  solution of the BVP of the form (\ref{14})--(\ref{18}) with $d_1(u) = a_1$, $d_2(v) = a_2$ and $\textbf{Q}(t) = \textbf{q}$ in the explicit form:
\begin{eqnarray}
& & u = \frac{u_v - u_m}{\Phi_1(R) - \Phi_1(\omega_2)} \Phi_1\left(\sqrt{x_1^2 + x_2^2 + (x_3 - \mu t)^2} + x_3 - \mu t\right) + \frac{u_m \Phi_1(R) - u_v \Phi_1(\omega_2)}{\Phi_1(R) - \Phi_1(\omega_2)},\nonumber \\ & & v = \frac{v_m - v_{\infty}}{\Phi_2(\omega_2)} \Phi_2\left(\sqrt{x_1^2 + x_2^2 + (x_3 - \mu t)^2} + x_3 - \mu t\right) + v_{\infty},\nonumber \\ & & S_k \equiv \frac{x_1^2 + x_2^2}{\omega_k^2} + \frac{2 (x_3 - \mu t)}{\omega_k} - 1 = 0, \ k = 1, 2; \ \omega_1 = R.\nonumber
\end{eqnarray}

\begin{remark} Whereas the basic equations of the BVP of the form (\ref{14})--(\ref{18}) with $d_1(u) = a_1$, $d_2(v) = a_2$ are linear, the relevant equations of  the initial BVP of the form (\ref{8})--(\ref{12}) may be nonlinear, but satisfying the conditions  $a_1 = \frac{\lambda_1(\phi^{-1}_1(u))}{C_1(\phi^{-1}_1(u))}$, $a_2 = \frac{\lambda_2(\phi^{-1}_2(v))}{C_2(\phi^{-1}_2(v))}$ (see the functions $\phi_1$ and $\phi_2$ in (\ref{13})).
\end{remark}

{\textbf{Example 2.}} \emph{The BVP of the form (\ref{4.9})--(\ref{4.13}) with $d_1(u) = u^{-1}$ and $d_2(v) = 1$, i.e., (\ref{4.9}) is the fast diffusion equation, while (\ref{4.10}) is the linear diffusion equation.}

In this case, the general solutions of equations (\ref{4.9}) and (\ref{4.10}) are given
 %in implicit form
  by the formulae \cite{pol-za}
\begin{equation}\label{4.16}
\int_{a}^{\omega u} \frac{d \nu}{\nu \left(1 + e^{- W\left(e^A\right) + A} \right)} = \ln \omega + C_2, \ v = C_3 \Phi(\omega) + C_4,
\end{equation}
where $\Phi(\omega) = \int_{\omega}^{+ \infty} \omega^{-1} e^{- \frac{\mu}{2} \omega} d\omega$, $W(x)$ is the Lambert function,
 $A = - \frac{\mu}{2} \nu + C_1$, $a$ is an arbitrary constant, $C_1, \ldots, C_4$ are to-be-determined constants.

Substituting solution (\ref{4.16}) into the boundary conditions (\ref{4.11})--(\ref{4.13}) and taking into account the formulae
\[
\frac{d \Phi}{d \omega} = \omega^{-1} e^{- \frac{\mu}{2} \omega}, \ \ln \left(\frac{\omega}{u} \frac{d u}{d \omega} \right) + \frac{\omega}{u} \frac{d u}{d \omega} = A,
\]
we obtain the  exact solution of the BVP in question:
\begin{equation}\label{4.17}
\int_{R u_v}^{\omega u} \frac{d \nu}{\nu \left(1 + e^{- W\left(e^{\mathcal{A}}\right) + \mathcal{A}} \right)} = \ln \frac{\omega}{R}, \ v = \frac{v_m - v_{\infty}}{\Phi(\omega_2)} \Phi(\omega) + v_{\infty},
\end{equation}
where the parameters $\omega_2$ and $\mu$ must be found from the system of transcendent equations
\begin{eqnarray}
& &\int_{R u_v}^{\omega_2 u_m} \frac{d \nu}{\nu \left(1 + e^{- W\left(e^{\mathcal{A}}\right) + \mathcal{A}} \right)} = \ln \frac{\omega_2}{R} ,\nonumber \\ & & 2 \frac{v_m - v_{\infty}}{\Phi(\omega_2)} e^{- \frac{\mu}{2} \omega_2} =  2 e^{- W\left(e^{\mathcal{A}(\omega_2)}\right) + \mathcal{A}(\omega_2)} + \mu \omega_2 H_m.\nonumber
\end{eqnarray}
Here, we used the designations
\[ \mathcal{A} = - \frac{\mu}{2} \nu + \ln \left(\left(\mu H_v - q \right) \frac{R}{2}\right) + \left(\mu H_v - q \right) \frac{R}{2} + \frac{\mu}{2} R u_v \]
\[ \mathcal{A}(\omega_2) =  -\frac{\mu}{2} \omega_2 u_m + \ln \left(\left(\mu H_v - q \right) \frac{R}{2} \right)+ \left(\mu H_v - q \right) \frac{R}{2}+ \frac{\mu}{2} R u_v .\]

Finally, using formulae (\ref{4.17}) , (\ref{4.2}) and (\ref{4.8}), we obtain the exact solution of the origin  BVP (\ref{14})--(\ref{18}) with $d_1(u) = u^{- 1}$, $d_2(v) = 1$ and $\textbf{Q}(t) = \textbf{q}$ in the implicit form
\begin{eqnarray}
& & \int_{R u_v}^{\left(\sqrt{x_1^2 + x_2^2 + (x_3 - \mu t)^2} + x_3 - \mu t \right) u} \frac{d \nu}{\nu \left(1 + e^{- W\left(e^{\mathcal{A}}\right) + \mathcal{A}} \right)} = \ln \frac{\sqrt{x_1^2 + x_2^2 + (x_3 - \mu t)^2} + x_3 - \mu t}{R},\nonumber \\ & & v = \frac{v_m - v_{\infty}}{\Phi(\omega_2)} \Phi\left(\sqrt{x_1^2 + x_2^2 + (x_3 - \mu t)^2} + x_3 - \mu t\right) + v_{\infty},\nonumber \\ & & S_k \equiv \frac{x_1^2 + x_2^2}{\omega_k^2} + \frac{2 (x_3 - \mu t)}{\omega_k} - 1 = 0, \ k = 1, 2; \ \omega_1 = R. \nonumber
\end{eqnarray}

It should be noted that several BVPs of the form (\ref{4.9})--(\ref{4.13}) can also be exactly solved for some other forms of diffusivity coefficients. Moreover, one may apply the standard program  package (e.g., Maple, Mathematica) to numerically solve   this BVP with arbitrary given diffusivities.

\section{\bf Conclusions}

In this paper, multi-dimensional nonlinear BVPs  with the basic evolution equations   by means of
the classical Lie symmetry method are studied. We consider BVPs  of the  most general form   (\ref{1})--(\ref{4}) , which include
basic equations of the arbitrary order, boundary conditions on  known and  unknown  moving
surfaces, boundary conditions on regular and non-regular manifolds.
A new
definition of invariance in the Lie sense for such  BVPs is formulated.  The definition  generalizes those proposed  earlier for  simpler BVPs \cite{ben-olv-82, b-k, ibr92,ch-kov-11}, and it can be
extended to BVPs  for  hyperbolic and elliptic  equations.
Note that  the comparison of this definition with those proposed earlier  is presented  in  the recent paper  \cite{ch-kov-11},  where Definition~1  in the    case of   (1+1)-dimensional BVP was formulated.
 %.
 In this paper, we   also propose the algorithm of the group classification for  classes of  BVPs, i.e., extending  the well-known problem for PDEs to BVPs.  Of course,  the group classification problem  for simple   classes of  BVPs can be solved in a straightforward way (see, e.g., \cite{ch-kov-09}), however, one needs  to determining  some  algorithm in the general case.

The main part of the paper is devoted to solving the group classification problem for  the class of (1+3)-dimensional  BVPs (\ref{8})--(\ref{12}), modeling processes of   melting and evaporation  under a powerful flux of energy. First of all, we simplified the BVPs  in question to the form  (\ref{14})--(\ref{18}) using the Goodman substitution. Since the basic equations of the problem obtained are the standard NHEs, we used the known system of  determining   equations  \cite{ch-king00,ch-king06} to derive their Lie symmetry description. Having done this and using the group of equivalence transformations, we proved  Theorem 2 presenting all possible Lie groups of  invariance of the BVPs of the form (\ref{14})--(\ref{18}).
 It was shown that  the Lie invariance does not  depend on the diffusivities $d_1(u)$, $d_2(v)$  but only in the form of the flux   $\textbf{Q}(t) $. There are only  five correctly specified forms of $\textbf{Q}(t) $  (see Table 2) leading to the extensions of the three-dimensional  Lie group of  invariance (the trivial  Lie  group), which is admitted by the arbitrary BVP of the form  (\ref{14})--(\ref{18}). One may note that cases  3 and 4  from Table 2 have no analogs in the (1+1)-dimensional  space of independent variables,  while  cases 5 and 6 are  generalizations of the relevant (1+1)-dimensional  BVPs (see \cite{ch-kov-09} for comparison).

We study in detail the BVP of the form (\ref{14})--(\ref{18}) with  arbitrary diffusivities and the special  form of the flux $\textbf{Q}(t) = \textbf{q}  $, which naturally arises as a mathematical model
of  the   melting and evaporation process. Since the MGI of this problem is five-dimensional,
the sets of  optimal  $s$-dimensional subalgebras were  constructed using the known algorithm \cite{pathera-1,bar}. The  brief analysis of these  subalgebras  is presented.   One of them, the two-dimensional  algebra,   is applied
for the reduction of the problem in question to
%Finally, the example how to reduce
%construct exact solution the (1+3)-dimensional nonlinear problem (\ref{14})--(\ref{18})
  the nonlinear  BVP for ODEs. Finally, the BVP obtained
%correctly-specified coefficients
 was exactly solved  in two cases of  correctly-specified  diffusivities;  hence,  the exact solutions   of the BVP of the form (\ref{14})--(\ref{18}) with  these   diffusivities were  found.
It should be noted that the BVP of the form (\ref{14})--(\ref{18}) with  constant  diffusivities was studied earlier in \cite{lyubov}, where the same result was obtained using  an ad hoc  ansatz, which does not connected with any Lie symmetry. To the  best of our  knowledge,   the exact solution of the BVP of the form (\ref{14})--(\ref{18}) with the fast diffusion constructed above   is new.

The work is in progress to apply the results obtained in this paper to the reduction and construction of  exact solutions for other
multi-dimensional BVPs with the remarkable Lie invariance.

\medskip
%\begin{center}
%\textbf{ Acknowledgment} \end{center}

%  The authors   are grateful
%to the     referees for the  very useful comments.

\end{document}